\documentclass[twocolumn,epj]{svjour}

\usepackage{amsmath,amssymb,amsfonts}
\usepackage{mathrsfs}
\usepackage{accents}

\usepackage{graphicx}
\usepackage{xcolor}
\usepackage{comment}

\usepackage{multirow}
\usepackage{booktabs}

\usepackage{algorithm}
\usepackage{algorithmicx}
\usepackage{algpseudocode}

\usepackage{textcomp}

\usepackage{manyfoot}

\usepackage{listings}

\usepackage[title]{appendix}

\usepackage[numbers,sort&compress]{natbib}

\usepackage[colorlinks,citecolor=blue,urlcolor=blue,linkcolor=blue]{hyperref}

\begin{document}
\pagenumbering{arabic}

\title{Correlated Systematic Uncertainties and Errors-on-Errors in Measurement Combinations with an Application to the $7$-$8$ TeV ATLAS-CMS Top Quark Mass Combination}

\author{Enzo Canonero \href{https://orcid.org/0000-0002-7180-4562}{\includegraphics[scale=0.06]{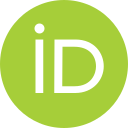}}
        \and
        Glen Cowan \href{https://orcid.org/0000-0001-8363-9827}{\includegraphics[scale=0.06]{orcid.png}}
}
\institute{Physics Department, Royal Holloway, University of London, U.K.}

\date{Received: date / Accepted: date}

\abstract{
The Gamma Variance Model is a statistical model that incorporates uncertainties in the assignment of systematic errors (informally called \textit{errors-on-errors}). The model is of particular use in analyses that combine the results of several measurements. In the past, combinations have been carried out using two alternative approaches: the Best Linear Unbiased Estimator (BLUE) method or what we will call the nuisance-parameter method. In this paper, we obtain a general relation between the BLUE and nuisance-parameter methods when the correlations induced by systematic uncertainties are non-trivial (i.e., not $\pm 1$ or $0$), and we then generalise the nuisance-parameter approach to include \textit{errors-on-errors}.  We then present analytical formulas for estimating central values, confidence intervals, and goodness-of-fit when \textit{errors-on-errors} are incorporated into the statistical model. To illustrate the properties of the Gamma Variance Model, we apply it to the $7$–$8$ TeV ATLAS-CMS top quark mass combination. We also explore a hypothetical scenario by artificially adding a fictitious measurement as an outlier to the combination, illustrating a key feature of the Gamma Variance Model — its sensitivity to the internal consistency of the input data — which could become relevant for future combinations.
}


\authorrunning{E.~Canonero and G.~Cowan}
\titlerunning{Correlated Systematic Uncertainties and Errors-on-Errors in Measurement Combinations}
\maketitle



\section{Introduction}
In Particle Physics analyses such as those at the Large Hadron Collider (LHC), many results are already dominated by systematic uncertainties or will be shortly.  In this context, uncertainties in the assigned values of systematic errors themselves -- informally called \textit{errors-on-errors} -- become increasingly important.  Often, systematic uncertainties are estimated using real datasets of control measurements or Monte Carlo simulations. In such cases, the limited size of these datasets introduces an uncertainty in the assigned value of the systematic error, which has a statistical nature. However, this is not the only source of uncertainty in its assigned value. For some systematic uncertainties, the definition of a ``$1$-$\sigma$'' error can itself be ambiguous. Theoretical uncertainties, for example, often rely on ad hoc estimation procedures that inherently carry an additional degree of uncertainty. Furthermore, ``two-point systematics'', where uncertainties are estimated from the difference in results from two methods, provide only a limited picture of the actual uncertainty. The Gamma Variance Model (GVM) \cite{bib:Cowan2019, bib:Canonero2023} offers a statistical framework to handle \textit{errors-on-errors} of both types.

In this paper, we focus on the application of the GVM to combinations, and we derive a number of new results that facilitate the use of the model in a non-trivial analysis. We first show how a likelihood with nuisance parameters can be constructed when the correlations induced by systematic uncertainties are not simply $\pm 1$ or $0$, and how this approach relates to the Best Linear Unbiased Estimator (BLUE) method~\cite{bib:Lyons, bib:Valassi2003, bib:Nisius2020, bib:Nisius2014}. Next, we demonstrate how this model can be generalized to include \textit{errors-on-errors} and derive analytical formulas for its application under these conditions. This framework is then applied to the $7$-$8$ TeV ATLAS-CMS top quark mass combination~\cite{bib:ATLAS/CMS}. Specifically, we investigate the impact on the combination's central value and confidence interval when the largest sources of systematic uncertainties are treated as uncertain. We present the results by considering the largest systematic uncertainties as uncertain, one at a time, and we vary their associated \textit{error-on-error} parameters. This approach serves as a general method to identify the systematic uncertainties to which a combination is sensitive when \textit{errors-on-errors} are considered. 

Finally, we explore how the inclusion of a hypothetical measurement that is in tension with the rest of the data affects the combination's sensitivity to uncertain systematics. This exercise can be relevant for future combinations, as some existing top-mass measurements deviate from the result of the ATLAS-CMS combined result~\cite{bib:Tevatron2014, bib:ATLAS2023}. This example illustrates the behavior of the GVM in scenarios where the inputs of a combination are not entirely internally compatible. Unlike a standard combination, incorporating \textit{errors-on-errors} makes the central value of the combination, here taken as the maximum likelihood estimator for the parameter of interest, more robust to discrepancies between input data values. Specifically, the central value becomes less sensitive to outliers. At the same time, the presence of discrepancies results in an increase in the confidence interval for the parameter of interest. This increase reflects the uncertainty stemming from the fact that the measurements in the dataset are not entirely compatible.


In Sec.~\ref{sec:BLUE}, we present an overview of how a combination can be performed using nuisance parameters under non-trivial correlation assumptions, and how this method relates to the BLUE approach. Section~\ref{sec:GVM} illustrates how to extend combinations to take \textit{errors-on-errors} into account. In Sec.~\ref{sec:useful-formulas}, we provide analytical formulas for profiling over nuisance parameters, and computing confidence intervals and goodness-of-fit when \textit{errors-on-errors} are incorporated into the statistical model. In Sec.~\ref{sec:top_mass}, we apply the GVM to the 7-8 TeV ATLAS-CMS top-mass combination and explore the impact of introducing an outlier to the combination. A summary and conclusions are presented in Sec.~\ref{sec:conclusions}.

\section{Equivalence between nuisance-parameter and BLUE methods in combinations with non-trivial correlations}

\label{sec:BLUE}
In Particle Physics, statistical data analysis aims at testing theoretical hypotheses $H$ using experimentally collected data, denoted as $\boldsymbol{y}$. The probability distribution of the data, $P(\boldsymbol{y}|H)$, is often indexed by a set of Parameters Of Interest (POIs) $\boldsymbol{\mu}$ and Nuisance Parameters (NPs) $\boldsymbol{\theta}$. NPs account for systematic uncertainties in the model, while the POIs are the main objective of the analysis. The likelihood function, $L(\boldsymbol{\mu}, \boldsymbol{\theta}) = P(\boldsymbol{y}|\boldsymbol{\mu}, \boldsymbol{\theta})$, is the central object needed for inference about the POIs. 

Here, we explore the application of combining  $N$  measurements $\boldsymbol{y} = (y_1, \ldots, y_N)$ of a parameter, each with its own statistical and systematic uncertainties, to obtain a single best estimate. We assume that \( \boldsymbol{y} \) follows a multivariate Gaussian distribution with expected values and covariance matrix given by

\begin{eqnarray}
    \label{eq:expecyi}
    \text{E}[y_i] & = &\mu+\sum_{s=1}^M\Gamma_i^s\theta_s \,, \\*[0.2 cm]
    \label{eq:covyiyj}
    V_{ij} & = & \text{cov}[y_i, y_j] \,,
\end{eqnarray}

\noindent where $M$ is the total number of NPs. Each NP $\theta_s$ represents a potential source of bias in the measured $y_i$ values, and the magnitude of these biases is described by the factors $\Gamma_i^s$.  Following this notation, we will use indices $i, j$, etc., to refer to components of $\boldsymbol{y}$ and $s$ to label a source of systematic uncertainty.

Additionally, independent \textit{control measurements} $\boldsymbol{u}$ are used to constrain the NPs. These control measurements $\boldsymbol{u} = (u_1,\ldots,u_M)$ are assumed to be best estimates of the NPs $\boldsymbol{\theta}=(\theta_1,\ldots,\theta_M)$ and are treated as independently Gaussian-distributed random variables with standard deviations $\boldsymbol{\sigma_u}= (\sigma_{u_1},\ldots,\sigma_{u_M})$. Under these assumptions, the log-likelihood of the model is 
\begin{equation}
\label{eq:comb_log_lik}
\begin{split}
    \ell(\mu,\boldsymbol{\theta})&=-\frac{1}{2}\sum_{i,j=1}^N\left(y_i-\mu-\sum_{s=1}^M\Gamma_i^{s}\theta_s\right) V_{ij}^{-1}\\ &\times\left(y_j-\mu-\sum_{s=1}^M\Gamma_j^{s}\theta_s\right)-\frac{1}{2}\sum_{s=1}^M\frac{(u_s-\theta_s)^2}{\sigma_{u_s}^2}\,.
\end{split}
\end{equation}

It is usually convenient to redefine the factors $\Gamma_i^s$, the NPs $\theta_{s}$ and the auxiliary measurements $u_s$ such that $u_s$ is Gaussian distributed about $\theta_s$ with a variance of unity, i.e., $\sigma_{u_s}^2 = 1$. With this choice, the value of \(\Gamma_i^s\) corresponds to the systematic error induced by source \(s\) in the measurement \(y_i\). Furthermore, in the real experiment, the best estimates of the $u_s$  are zero.  The idea here is that if there are known biases in the measurements, these are already subtracted. The remaining systematic uncertainty lies in the fact that this bias removal might be imprecise, due to the limited accuracy with which we can identify and quantify the biases.
Nevertheless, here we will retain the complete notation in our discussion for greater generality. Auxiliary measurements must be explicitly included to perform Monte Carlo (MC) simulations, while the systematic variances  $\sigma^2_{u_s}$ are needed to generalise the model to include errors-on-errors.

The above likelihood can be related to the Best Linear Unbiased Estimate (BLUE) method \cite{bib:Lyons, bib:Valassi2003, bib:Nisius2020, bib:Nisius2014}, which is a commonly used technique for performing combinations that does not require the introduction of NPs. This method involves constructing an approximate Gaussian likelihood to describe the $N$ measurements of the combination, $\boldsymbol{y}$, based on a covariance matrix $W_{ij}$ that includes both statistical and systematic uncertainties. The resulting log-likelihood

\begin{equation}
\label{eq:blue}
    \ell(\mu) = -\frac{1}{2}\sum_{i,j=1}^N (y_i-\mu) \, W^{-1}_{ij} \,(y_j-\mu)
\end{equation}

\noindent is then maximized to get the best estimate of $\mu$. The fundamental assumption of the method is that the covariance matrix $W$ can be expressed as a sum of terms: one for the statistical uncertainties $V$ and one for each systematic source $U^{(s)}$:

\begin{equation}
\label{eq:V_blue}
    W_{ij} =  V_{ij} + \sum_{s=1}^M U^{(s)}_{ij}\,.
\end{equation}

\noindent Here the index $s$ runs over each source of systematic uncertainty. These systematics are independent from each other in the sense that their associated control measurements are uncorrelated, but they induce correlations between the $\boldsymbol{y}$ measurements. This underlying structure is more evident when Eq.\eqref{eq:comb_log_lik} is used instead. In fact, the two methods are equivalent when the auxiliary measurements are set to zero, $u_s = 0$. Specifically, by profiling Eq.\eqref{eq:comb_log_lik} over all the NPs, one can recover the BLUE log-likelihood as defined in Eq.\eqref{eq:blue}.\footnote{Profiling over a parameter means maximizing the likelihood with respect to that parameter while keeping all others fixed.} Furthermore, the relationship between the factors $\Gamma_i^s$ and the BLUE covariance matrix $W$ uses Eq.~\eqref{eq:V_blue} with

\begin{equation}
\label{eq:blue_cov_s}
    U^{(s)}_{ij}= \Gamma_i^s\,\Gamma_j^s\,\sigma_{u_s}^2\,.
\end{equation}
The equivalence between these two methods is widely discussed in the literature; more details can be found, for example, in~\cite{bib:Pinto2024, bib:Demortier1999, bib:Fogli2002, bib:Stump2001, bib:Thorne2002, bib:Botje2002, bib:Glazov2005, bib:Barlow2021, bib:List2010, bib:Aad2014}.

\subsection{Systematics with non-trivial correlations}
A limitation of the NP approach is that each systematic effect \( s \) can only induce correlation coefficients of \( \pm 1 \) or \( 0 \), which we call "trivial" correlations:
\begin{equation}
    \rho^{(s)}_{ij} = \frac{U^{(s)}_{ij}}{\sqrt{U^{(s)}_{ii}U^{(s)}_{jj}}} = \pm 1 \text{ or } 0\,.
\end{equation}
This restriction arises from the definition of the systematic terms $U^{(s)}_{ij}$ in the BLUE covariance matrix, as specified in Eq.~\eqref{eq:blue_cov_s}. Generalising the BLUE method to accommodate this is very simple as it only requires modifying Eq.~\eqref{eq:blue_cov_s} to be

\begin{equation}
\label{eq:blue_cov_extened}
    U^{(s)}_{ij}= \rho^{(s)}_{ij} \Gamma_i^s \Gamma_j^s \sigma_{u_s}^2\,. 
\end{equation}
Note that one may not, however, assign arbitrary values to the $\rho^{(s)}_{ij}$ as this could lead to $U^{(s)}$ not being positive definite.  

It is conceptually more complicated to extend the log-likelihood containing NPs of Eq.~\eqref{eq:comb_log_lik} so that it gives non-trivial correlations. To accomplish this, one needs to define a distinct NP \(\theta^i_s\) associated with the systematic effect \(s\) for each measurement \(y_i\), and introduce correlations between the corresponding auxiliary measurements such that
\begin{equation}
\label{eq:u_cov}
    \mbox{cov}[u^i_s, u^j_s] = \sigma_{u_s}^2 \rho^{(s)}_{ij}\,.
\end{equation}
Thus the log-likelihood of Eq.~\eqref{eq:comb_log_lik} becomes
\begin{equation}
\label{eq:comb_log_lik_u}
\begin{split}
    \ell(\mu,\boldsymbol{\theta})&=-\frac{1}{2}\sum_{i,j=1}^N\left(y_i-\mu-\sum_{s=1}^M\Gamma_i^{s}\theta^i_s\right) V_{ij}^{-1}\\ &\times\left(y_j-\mu-\sum_{s=1}^M\Gamma_j^{s}\theta^j_s\right)\\&-\frac{1}{2}\sum_{s=1}^M\sum_{i,j=1}^N(u^i_s-\theta^i_s)\frac{1}{\sigma_{u_s}^2}\left(\rho^{(s)}\right)_{ij}^{-1}(u^j_s-\theta^j_s)\,.
\end{split}
\end{equation}

\noindent In App.~\ref{app:A} we prove the important result that this log-likelihood becomes equivalent to the BLUE method when the auxiliary measurements $u^i_s$ are set to zero. This equivalence can be demonstrated by profiling Eq.\eqref{eq:comb_log_lik_u} with respect to all the NPs, which leads to the BLUE likelihood as defined in Eq.\eqref{eq:blue}, with covariance matrix specified by Eqs.~\eqref{eq:V_blue} and~\eqref{eq:blue_cov_extened}. 

\subsection{Comparison with the Convino framework}
Correlations between auxiliary measurements in combinations are implemented within the software package proposed in~\cite{bib:Kieseler2017} called Convino. In this framework, auxiliary measurements associated with different sources of systematic uncertainties can be treated as correlated. This approach leads to a systematic constraint term, expressed in our notation as
\begin{equation}
    -\frac{1}{2}\sum_{s=1}^M (u_s-\theta_s)C_{sp}^{-1}(u_p-\theta_p)\,,
\end{equation}
where \(C\) is the covariance matrix representing the correlations among auxiliary measurements associated with different systematic effects. This matrix is crucial for handling fits with simultaneously constrained uncertainties, where the NP estimates are correlated.

In contrast, our approach focuses on extending the NP basis for each individual systematic source \(s\) to induce the desired correlation structure in the measurements \(\boldsymbol{y}\). However, we do not introduce correlations between different sources of systematic uncertainties, i.e., auxiliary measurements with different \(s\) indices remain independent. Specifically, for each systematic \(s\), we introduce a separate NP \(\theta_s^i\) and auxiliary measurement \(u_s^i\) for each measurement \(i\).  This leads to correlations only between auxiliary measurements that share the same \(s\) index but have different \(i\) indices, and results in a systematic constraint term given by 
\begin{eqnarray}
    -\frac{1}{2}\sum_{s=1}^M\sum_{i,j=1}^N(u^i_s-\theta^i_s)\frac{1}{\sigma_{u_s}^2}\left(\rho^{(s)}\right)_{ij}^{-1}(u^j_s-\theta^j_s)\,.
\end{eqnarray}
This extension is necessary to achieve complete equivalence between the BLUE method and the nuisance-parameter approach under any correlation assumption.

After introducing new NPs \(\theta_s^i\) for a given source \(s\), the Convino framework offers a natural, and more general, extension by allowing correlations between different systematic effects. More specifically, it allows for correlations among auxiliary measurements that have distinct \(s\) indices. Nevertheless, by mapping \(s\) and \(i\) onto a single index using a mapping function \(f(s, i)\), the Convino framework can be applied in its current form to incorporate the NP approach discussed here. However, the formalism we use clearly highlights the connection between different independent systematic sources and the correlation structure they induce in the measurements \(\boldsymbol{y}\),  as defined within the framework of the BLUE method.

\section{Extending combinations to account for \textit{errors-on-errors}}
\label{sec:GVM}

In this section, we demonstrate how to extend a combination to account for the possibility that a systematic uncertainty may itself be uncertain. These uncertainties, referred to as \textit{errors-on-errors}, can arise from ambiguities in the definition of a $1-\sigma$ error, as is often the case with theoretical uncertainties or two-point systematics. Additionally, they can result from the estimation of a systematic effect based on a limited sample of MC events or real data. Accounting for such uncertainties can have a non-negligible, or even significant, impact on the combination results, as demonstrated in the example provided in Sec.~\ref{sec:top_mass}.

\subsection{The GVM with trivial correlations}
\label{sec:GVMorig}
The likelihood of Eq.~\eqref{eq:comb_log_lik} can be extended to account for the presence of uncertainties in the assignment of systematic errors, informally denoted as \textit{errors-on-errors}. This is achieved using the Gamma Variance Model (GVM)~\cite{bib:Cowan2019, bib:Canonero2023}. This model allows the systematic uncertainty \(\sigma_{u_s}\) to be treated as uncertain. Specifically, its square, the variance \(\sigma_{u_s}^2\), is treated as an adjustable parameter of the model, effectively introducing it as an NP. For mathematical convenience, we use the variance \(\sigma^2_{u_s}\) as the model's NP rather than the standard deviation \(\sigma_{u_s}\). In the GVM, the value that would typically be assigned to \(\sigma^2_{u_s}\) in a standard combination is treated as an estimate of this parameter and is denoted as \(v_s\).

One should keep in mind that the choice of independent gamma distributions for the estimates \(\boldsymbol{v}\) is not expected to produce a perfect model, rather an approximation that represents a potentially significant improvement over the assumption of exactly specified variances.  Other distributions, provided they are defined for positive variables, could also be considered.  Some studies carried out in connection with \cite{bib:Cowan2019} of the log-normal distribution or a gamma distribution for the estimated standard deviations led to qualitatively similar results, but were not pursued further when the following reasons to use gamma-distributed \(\boldsymbol{v}\) became apparent.  

First, suppose that $u_s$ itself is found as an average of some number of other Gaussian-distributed values $u_{s,i}$, $i = 1,...,n$, and the estimate of $\sigma^2_{u_s}$ is found as the sample variance of these $n$ values.  In this case, the estimated variance follows a gamma distribution.  In general, the $u_s$ might not be found in this way and the gamma distribution therefore is not expected to provide an exact model.  The scenario of multiple Gaussian measurements leading to $u_s$ is nevertheless plausible and provides an important point of contact with our model assumptions.

Second, the gamma distribution provides significant mathematical simplifications, as will be demonstrated in this section. Detailed methodology and motivation for this approach can be found in~\cite{bib:Cowan2019}. 

The log-likelihood of GVM is defined as
\begin{equation}
\label{eq:gvm_avareges_full}
\begin{split}
\ell(\mu,\boldsymbol{\theta}, \boldsymbol{\sigma_{u}^2})&=-\frac{1}{2}\sum_{i,j=1}^N\left(y_i-\mu-\sum_{s=1}^M\Gamma_i^{s}\theta_s\right) V_{ij}^{-1}\\ &\times \left(y_j-\mu-\sum_{s=1}^M\Gamma_j^{s}\theta_s\right)-\frac{1}{2}\sum_{s=1}^M\frac{(u_s-\theta_s)^2}{\sigma^2_{u_s}}\\&-\frac{1}{2}\sum_{s=1}^M\bigg[\left(1+\frac{1}{2\varepsilon^2_s}\right)\log{\sigma^2_{u_s}}+\frac{v_s}{2\varepsilon_s^2\sigma^2_{u_{s}}}\bigg]\,.
\end{split}
\end{equation}
This model extends Eq.~\eqref{eq:comb_log_lik} by adding, in the last line, terms associated with treating \(\sigma^2_{u_s}\) as an adjustable parameter of the model and its estimate \(v_s\) as gamma distributed. The parameter $\varepsilon_s$ quantifies the relative uncertainty associated with the systematic error $\sigma_{u_s}$ and is commonly referred to as the \textit{error-on-error} parameter. For instance, if $\varepsilon_s$ equals $0.3$, this indicates that $s_s = \sqrt{v_s}$ has a 30\% relative uncertainty as an estimate of $\sigma_{u_s}$. 

In principle, the \(\varepsilon_s\) parameters should be provided by the analyst based on any available information regarding the uncertainties in \(\sigma_{u_s}\). It can represent the statistical uncertainty of \(\sigma_{u_s}\) arising from the limited size of the auxiliary measurement dataset or be assigned using more ad hoc arguments for uncertainties such as theoretical errors. In the absence of the required expert knowledge to inform such a choice, results of the GVM may be presented as a function of assumed values for $\varepsilon_s$. 

One of the appealing mathematical properties of this model is that the log-likelihood can be maximized in closed form with respect to \(\sigma^2_{u_s}\), for fixed \(\mu\) and \(\boldsymbol{\theta}\), using
\begin{equation}
\label{eq:sigma_mle1}   \widehat{\widehat{\sigma_{u_s}^2}} = \frac{v_s+2\varepsilon_s(u_s-\theta_s)^2}{1+2\varepsilon_s^2}\,.
\end{equation}
This yields the profile log-likelihood $\ell_p(\mu, \boldsymbol{\theta}) = \ell(\mu, \boldsymbol{\theta}, \boldsymbol{\widehat{\widehat{\sigma_{u}^2}}})$ 
\begin{equation}
\label{eq:gvm_avareges}
\begin{split}
\ell_p(\mu,\boldsymbol{\theta})&=-\frac{1}{2}\sum_{i,j=1}^N\left(y_i-\mu-\sum_{s=1}^M\Gamma_i^{s}\theta_s\right) V_{ij}^{-1}\\ &\times \left(y_j-\mu-\sum_{s=1}^M\Gamma_j^{s}\theta_s\right)\\&-\frac{1}{2}\sum_{s=1}^M\left(1+\frac{1}{2\varepsilon_s^2}\right)\log\left[1+2\varepsilon_s^2\frac{(u_s-\theta_s)^2}{v_s}\right]\,.
\end{split}
\end{equation}
In this discussion, we denote \(\sigma_{u_s}\) - the standard deviation of \(u_s\) - as the systematic error. However, in a manner similar to that of Sec.~\ref{sec:BLUE}, it is often convenient to redefine \(\Gamma^s_i\), \(\theta_s\), and \(u_s\) so that, in the absence of \textit{errors-on-errors} (i.e., when \(\sigma_{u_s}\) is precisely known), the variable \(u_s\) is Gaussian-distributed with mean \(\theta_s\) and unit variance. Consequently, the best estimates of all variances become \(v_s = 1\). Under this choice, the actual estimate of the systematic error induced by source \(s\) in measurement \(y_i\) is expressed as \(\Gamma_i^s\). Thus, one simply needs to fix the \(\Gamma_i^s\) factors to the systematic uncertainty values. This redefinition does not affect the method of introducing \textit{errors-on-errors}, as the model now accounts for potential fluctuations in \(v_s\), which rescale \(\Gamma_i^s\) whenever the systematic error is over- or underestimated.

The inclusion of errors-on-errors replaces the standard quadratic constraints of Eq.~\eqref{eq:comb_log_lik}, typically used to model systematic effects, with the logarithmic constraints of Eq.~\eqref{eq:gvm_avareges}. These logarithmic terms are quadratic and identical to those in Eq.~\eqref{eq:comb_log_lik} when \(\varepsilon_s\) approaches zero. However, for larger values of \(\varepsilon_s\), these log-terms penalize the log-likelihood less for values of \(\theta_s\) that deviate from \(u_s\), thus changing the way the model handles incompatible data. An example of this property will be provided in Sec.~\eqref{sec:outliers}. An application of this model has also been presented in~\cite{bib:Reader2024} in the context of Parton Distribution Function (PDF) fitting.

\subsection{The GVM with non-trivial correlations}
\label{sec:GVMcorr}
The likelihood of Eq.\eqref{eq:comb_log_lik_u} is designed to admit non-trivial correlations, i.e., not only $0$ or $\pm1$. This is achieved by extending the NP basis for each systematic source \(s\), introducing a distinct NP \(\theta_s^i\) and auxiliary measurement \(u_s^i\) for each measurement \(i\), with correlations applied only among auxiliary measurements sharing the same \(s\) index but different \(i\) indices. 

Generalising it to include \textit{errors-on-errors} requires extending the GVM beyond what was originally presented in~\cite{bib:Cowan2019}. The method to achieve this remains similar to the previously discussed approach: the variances \(\sigma_{u_s}^2\) in Eq.~\eqref{eq:comb_log_lik_u} are treated as adjustable parameters within the model, meaning their exact values are now considered unknown. Meanwhile, their best estimates, \(v_s\), are modeled as independently gamma-distributed random variables and incorporated into the model. The resulting log-likelihood is
\begin{equation}
\label{eq:prof_comb_log_lik_u_full}
\begin{split}
    \ell(\mu,\boldsymbol{\theta}, \boldsymbol{\sigma_{u}^2})&=-\frac{1}{2}\sum_{i,j=1}^N\left(y_i-\mu-\sum_{s=1}^M\Gamma_i^{s}\theta^i_s\right)V_{ij}^{-1}\\&\times \left(y_j-\mu-\sum_{s=1}^M\Gamma_j^{s}\theta^j_s\right)\\&-\frac{1}{2}\sum_{s=1}^M\sum_{i,j=1}^N(u^i_s-\theta^i_s)\frac{1}{\sigma_{u_s}^2}\left(\rho^{(s)}\right)_{ij}^{-1}(u^j_s-\theta^j_s)\\&-\frac{1}{2}\sum_{s=1}^M\left[\left(N+\frac{1}{2\varepsilon^2_s}\right)\log{\sigma^2_{u_s}}+\frac{v_s}{2\varepsilon_s^2\sigma^2_{u_{s}}}\right]\,.
\end{split}
\end{equation}
Here, the terms in the last line arise from treating \(\sigma^2_{u_s}\) as an adjustable parameter of the model and \(v_s\), its estimate, as gamma-distributed. This log-likelihood can be maximized with respect to $\sigma^2_{u_s}$ using
\begin{equation}
\label{eq:sigma_mle2}   \widehat{\widehat{\sigma_{u_s}^2}}=\frac{v_s+2\varepsilon_s^2\sum_{i,j}^N(u_s^i-\theta_s^i)\left(\rho^{(s)}\right)_{ij}^{-1}(u_s^j-\theta_s^j)}{1+2N\varepsilon_s^2}\,,
\end{equation}
to obtain the profile log-likelihood:
\begin{equation}
\label{eq:prof_comb_log_lik_u}
\begin{split}
    \ell_p(\mu,\boldsymbol{\theta})&=-\frac{1}{2}\sum_{i,j=1}^N\left(y_i-\mu-\sum_{s=1}^M\Gamma_i^{s}\theta^i_s\right) V_{ij}^{-1}\\ &\times \left(y_j-\mu-\sum_{s=1}^M\Gamma_j^{s}\theta^j_s\right)\\&-\frac{1}{2}\sum_{s=1}^M \left(N+\frac{1}{2\varepsilon^2_s}\right) \log \left[1 + \frac{2\varepsilon_s^2}{v_s} \sum_{i,j=1}^N(u^i_s-\theta^i_s)\right.\\& \left.\times\left(\rho^{(s)}\right)_{ij}^{-1}(u^j_s-\theta^j_s)\right]\,.
\end{split}
\end{equation}
In this model, one would typically redefine the parameters in such a way that the estimate \(v_s\) is fixed to one. For a given source \(s\), any overestimation or underestimation of the systematic uncertainties induced in the measurements \(\boldsymbol{y}\) occurs uniformly by the same factor.

\subsection{Relationship between the two GVM versions}
In this section, we introduced two methods for incorporating errors-on-errors. To clarify their differences, consider a simplified example with two systematic effects, \(A\) and \(B\). In the first approach, the auxiliary measurements \(u_A\) and \(u_B\) have variances \(\sigma^2_{u_A}\) and \(\sigma^2_{u_B}\), with their best estimates \(v_A\) and \(v_B\) treated as independent random variables. This means the statistical fluctuations of \(v_A\) and \(v_B\) are uncorrelated, allowing scenarios where \(v_A\) underestimates \(\sigma^2_{u_A}\) while \(v_B\) overestimates \(\sigma^2_{u_B}\), or vice versa, or where both over- or underestimate their respective variances. These assumptions result in the following logarithmic constraints:
\begin{equation}
\begin{split}
&- \frac{1}{2} \left( 1 + \frac{1}{2\varepsilon_A^2} \right) \log \left[ 1 + 2\varepsilon_A^2 \frac{(u_A - \theta_A)^2}{v_A} \right] \\
& - \frac{1}{2} \left( 1 + \frac{1}{2\varepsilon_B^2} \right) \log \left[ 1 + 2\varepsilon_B^2 \frac{(u_B - \theta_B)^2}{v_B} \right]\,,
\end{split}
\end{equation}
where \(\varepsilon_A\) and \(\varepsilon_B\) are the \textit{errors-on-errors} parameters. This construction defines the model in Eq.~\eqref{eq:gvm_avareges} that we use to handle trivial correlations.

In the second approach, \(v_A\) and \(v_B\) are assumed to be fully dependent, meaning they either both underestimate or both overestimate their respective variances by the same factor. This is modeled by setting \(\sigma^2_{u_A} = \sigma^2 \Gamma^2_A\) and \(\sigma^2_{u_B} = \sigma^2 \Gamma^2_B\), where \(\sigma^2\) is an unknown global variance parameter, shared by both systematic effects. The best estimate of \(\sigma^2\) is denoted by \(v\), with an associated error-on-error parameter \(\varepsilon\). This approach leads to the following logarithmic term:
\begin{equation}
\small 
- \frac{1}{2} \left( N + \frac{1}{2\varepsilon^2} \right) \log \left[ 1 + 2\varepsilon^2 \frac{(u_A - \theta_A)^2}{v\Gamma^2_A} +  2\varepsilon^2 \frac{(u_B - \theta_B)^2}{v\Gamma_B^2} \right]\,,
\end{equation}
where \(N=2\), as the logarithm includes terms corresponding to two auxiliary measurements. The scaling factors \(\Gamma^2_A\) and \(\Gamma^2_B\) can be absorbed into the definitions of the NPs and auxiliary measurements for simplicity. Importantly, when \(\varepsilon = 0\), this approach becomes equivalent to the first method. Additionally, this method enables the introduction of a correlation factor between the auxiliary measurements \(u_A\) and \(u_B\), while also accounting for a global \textit{error-on-error} affecting the scale of their variances.

The model defined in Eq.~\eqref{eq:prof_comb_log_lik_u} to handle non-trivial correlations incorporates both assumptions. It treats the variances of auxiliary measurements from different systematic sources (indexed by \(s\)) as independent. In contrast, for a given source \(s\), the variances of auxiliary measurements (indexed by \(i\)) are considered dependent, as they are associated with the same systematic effect. That is, the estimates of the variances associated with a given systematic effect \( s \) are proportional to \( v_s \).

\section{Useful formulas for practical implementation of \textit{errors-on-errors}}
\label{sec:useful-formulas}
In this section, we present a set of analytical formulas designed to simplify the practical implementation of uncertain systematic errors. While the details are somewhat technical, they provide solutions to challenges that might otherwise complicate the application of the GVM. Specifically, the logarithmic constraints in Eqs.~\eqref{eq:gvm_avareges} and~\eqref{eq:prof_comb_log_lik_u} prevent exact analytical maximization of the GVM log-likelihoods with respect to \(\mu\) and \(\boldsymbol{\theta}\), as well as the computation of confidence intervals for \(\mu\) and the evaluation of goodness-of-fit for the observed data. Although these tasks can always be performed using numerical techniques, analytical formulas are particularly useful for reducing the computational burden that can arise in complex analyses. Therefore, they should be employed when the analyst believes they offer a clear advantage over numerical methods. Here, we provide approximate analytical solutions in the form of perturbative expansions in the error-on-error parameters \(\varepsilon_s^2\).

In this section, we focus on the GVM with trivial correlations. The case of the GVM with non-trivial correlations is addressed in App.~\ref{app:B}.

\subsection{Profiled values of nuisance parameters}
\label{sec:central_values}
In statistical modeling, NPs are introduced to account for systematic uncertainties. To reduce the dimensionality of the parameter space and facilitate the optimization of the likelihood, it is common to compute the profiled values of the NPs. These are the values of the NPs that maximize the likelihood for a fixed value of $\mu$. We denote such profiled values as \(\hat{\hat{\theta}}_s\).

To compute the profiled values of the NPs, we maximize the log-likelihood defined in Eq.~\eqref{eq:gvm_avareges} with respect to $\boldsymbol{\theta}$, while treating $\mu$ as fixed. This is done by solving the score equations \(\partial \ell_p / \partial \theta_s = 0\), leading to

\begin{equation}
\begin{split}
\label{eq:score_equations}
\sum_{i,j=1}^N &\Gamma_i^{s} V_{ij}^{-1} \left( y_j - \mu - \sum_{p=1}^M \Gamma_j^{p} \hat{\hat{\theta}}_p \right)\\ &+ \frac{1 + 2\varepsilon_s^2}{v_s + 2\varepsilon_s^2 \left( u_s - \hat{\hat{\theta}}_s \right)^2} \left( u_s - \hat{\hat{\theta}}_s \right) = 0\,.
\end{split}
\end{equation}

The score equations for the NPs form a system of coupled cubic equations, which do not have a general analytical solution. While it is always possible to perform a numerical optimization, it is also possible to solve them perturbatively at all orders in $\varepsilon_s^2$, using a recursive approach. We express $\hat{\hat{\theta}}_s$ as:
\begin{equation}
\label{eq:expansion1}
\hat{\hat{\theta}}_s = \hat{\hat{\theta}}_s^{(0)} + \varepsilon_s^2 \hat{\hat{\theta}}_s^{(1)} + \varepsilon_s^4 \hat{\hat{\theta}}_s^{(2)} + \cdots\,,
\end{equation}
where $\hat{\hat{\theta}}_s^{(0)}$ is the solution when $\varepsilon_s^2 = 0$, corresponding to the case without \textit{errors-on-errors}. Substituting this expansion into Eq.~\eqref{eq:score_equations} and equating terms of equal powers of $\varepsilon_s^2$, we obtain a recursive set of equations for the coefficients $\hat{\hat{\theta}}_s^{(n)}$. The first term is given by
\begin{equation} \hat{\hat{\theta}}_s^{(0)} = \sum_{p=1}^M \left(C^{(0)}\right)_{sp}^{-1} \left[ \sum_{i,j=1}^N \Gamma_i^{p} V_{ij}^{-1} \left(y_j - \mu\right) + \frac{u_p}{v_p} \right]\,, \end{equation}
where the matrix $C^{(0)}_{sp}$ is defined as
\begin{equation} 
C^{(0)}_{sp} = \sum_{i,j=1}^N \Gamma_i^{s} V_{ij}^{-1} \Gamma_j^{p} + \frac{\delta_{sp}}{v_s} \,
\end{equation}
and $\delta_{sp}$ is the Kronecker delta. Notice that this result is also the solution to the quadratic likelihood of Eq.~\eqref{eq:comb_log_lik}. At a generic order $\varepsilon_s^{2n}$, with $n\geq 1$, the perturbative factor $\varepsilon_s^{2n} \hat{\hat{\theta}}_s^{(n)}$ is

\begin{equation} 
\begin{split}
\varepsilon_s^{2n} \hat{\hat{\theta}}_s^{(n)} &= \sum_{p=1}^M \left(C^{(n)}\right)_{sp}^{-1}\left[\sum_{i,j=1}^N \Gamma_i^{p} V_{ij}^{-1} \bigg(y_j - \mu \right.\\ & \left.-\sum_{p'=1}^M \Gamma_j^{p'}T_{p',n-1}\bigg) + \frac{u_p - T_{p,n-1}}{S_{p,n-1}^{2}} \right]\,, 
\end{split}
\end{equation}
where the matrix $C^{(n)}_{sp}$ is defined as
\begin{equation} 
C^{(n)}_{sp} = \sum_{i,j=1}^N \Gamma_i^{s} V_{ij}^{-1} \Gamma_j^{p} + \frac{\delta_{sp}}{S_{s,n-1}^{2}} \,, \end{equation}
while \(T_{s,n}\) and \(S_{s,n}^{2}\) are given by
\begin{equation}
   T_{s,n} = \hat{\hat{\theta}}_s^{(0)} + \varepsilon_s^2 \hat{\hat{\theta}}_s^{(1)} + \cdots + \varepsilon_s^{2n} \hat{\hat{\theta}}_s^{(n)}\,,
\end{equation}
\begin{equation}
    S_{s,n}^{2} = \frac{v_s + 2\varepsilon_s^2 \left(u_s - T_{s,n}\right)^2}{1 + 2\varepsilon_s^2} \,.
\end{equation}
The variables \(T_{s,n}\) and \(S_{s,n}^{2}\) represent the order \(\varepsilon_s^{2n}\) approximations of \(\hat{\hat{\theta}}_s\) and \(\widehat{\widehat{\sigma_{u_s}^2}}\), respectively.

The convergence of the perturbative expansion requires the condition
\begin{equation}
\label{eq:condition1}
\frac{2\varepsilon_s^2}{v_s}(u_s - \hat{\hat{\theta}}_s)^2 < 1
\end{equation}
to hold, defining the radius of convergence for the series. This condition is violated either when the \textit{errors-on-errors} parameters are large or when the profiled values of the NPs deviate significantly from their pre-fit estimates (i.e., the associated auxiliary measurements).

When this condition is not satisfied, the approximate solutions provided in this section are not guaranteed to hold. In practice, we observe that these approximations break down when Eq.~\eqref{eq:score_equations} admits three real solutions instead of one, and the absolute maximum of the log-likelihood discontinuously shifts between different maxima. The more NPs fail to satisfy this condition, the more likely it is to encounter this scenario. However, when this occurs, it does not mean that the GVM is inapplicable; rather, it suggests that numerical maximization should be used instead.

Using the perturbative solutions for \(\hat{\hat{\theta}}_s\), the log-likelihood can be optimized with respect to \(\mu\) to determine its best estimate.

\subsection{Confidence intervals}
\label{sec:CI}
In frequentist statistics, a confidence region in the full parameter space \( (\mu, \boldsymbol{\theta}) \) can be constructed by testing hypothesized parameter values using the likelihood-ratio test statistic,
\[
\label{eq:lik_ratio}
    w_{\mu \boldsymbol{\theta}} = -2\log\left(\frac{L(\mu, \boldsymbol{\theta})}{L(\hat{\mu}, \hat{\boldsymbol{\theta}})}\right) = 2\left[\ell(\hat{\mu}, \hat{\boldsymbol{\theta}}) - \ell(\mu, \boldsymbol{\theta})\right] \,.
\]
When constructing confidence intervals for \( \mu \) only, we use the \emph{profile} likelihood ratio:
\[
\label{eq:prof_lik_ratio}
    w_{\mu} = -2\log\left(\frac{L(\mu, \hat{\hat{\boldsymbol{\theta}}})}{L(\hat{\mu}, \hat{\boldsymbol{\theta}})}\right) = 2\left[\ell(\hat{\mu}, \hat{\boldsymbol{\theta}}) - \ell(\mu, \hat{\hat{\boldsymbol{\theta}}})\right].
\]
Here, \(\hat{\mu}\) and \(\hat{\boldsymbol{\theta}}\) are the Maximum Likelihood Estimators (MLEs), which are the values of $\mu$ and $\boldsymbol{\theta}$ that maximize the likelihood, while \(\hat{\hat{\boldsymbol{\theta}}}\) are the profiled values of $\boldsymbol{\theta}$, obtained by maximizing the likelihood with \(\mu\) fixed.

Confidence intervals are typically computed under the assumption that the likelihood ratio \( w_{\mu \boldsymbol{\theta}} \) asymptotically follows a \(\chi^2\) distribution with \( 1 + M \) degrees of freedom, where \( M \) is the number of NPs in the model. This holds in the large sample (also called \textit{asymptotic}) limit (see, e.g., \cite{bib:Cowan2011, bib:Algeri2020}).  In this limit, the profile likelihood ratio \( w_{\mu} \) is \(\chi^2\)-distributed with 1 degree of freedom, corresponding to the single POI. This results in two equivalent methods for estimating confidence intervals: either by imposing \( w_{\mu} < 1 \) or by requiring \( \ell(\hat{\mu}, \hat{\boldsymbol{\theta}}) - \ell(\mu, \hat{\hat{\boldsymbol{\theta}}}) < \frac{1}{2} \).

For Gaussian data with linear dependence on model parameters, these statistics are inherently \(\chi^2\)-distributed, making the asymptotic limit exact for the Gaussian models defined by Eqs.~\eqref{eq:comb_log_lik} and~\eqref{eq:comb_log_lik_u}. However, the GVM\\ log-likelihoods in Eqs.~\eqref{eq:gvm_avareges_full} and~\eqref{eq:prof_comb_log_lik_u_full} contain non-quadratic terms, causing deviations from the \(\chi^2\) distribution by terms of order \( \varepsilon_s^2 \), which affect the precision of confidence intervals. 

We address this issue using the Bartlett correction~\cite{bib:Bartlett1937, bib:Codeiro2014}. The Bartlett correction belongs to a part of statistics called \textit{higher-order asymptotics}~\cite{bib:Brazzale2007}, which addresses deviations from asymptotic limits in statistical models. A discussion on how to apply \textit{higher-order asymptotics} to the GVM can be found in~\cite{bib:Canonero2023}. 

The idea behind the Bartlett correction is to adjust the likelihood ratio by a scaling factor so that its distribution more closely matches a \(\chi^2\) distribution, reducing error terms to order \(\varepsilon_s^4\), while also ensuring that \(p\)-values remain accurate up to terms of the same order. Specifically, the Bartlett-corrected likelihood ratio is defined as
\begin{equation}
\label{eq:wstarDef}
    w_{\mu \boldsymbol{\theta}}^\ast = w_{\mu \boldsymbol{\theta}} \,\left(\frac{1+M}{\text{E}[w_{\mu \boldsymbol{\theta}}]}\right)\equiv  \frac{w_{\mu \boldsymbol{\theta}}}{1 + b_{\mu \boldsymbol{\theta}}/(1+M)}\,,
\end{equation}
where $b_{\mu \boldsymbol{\theta}} = \text{E}[w_{\mu \boldsymbol{\theta}}] - 1 - M$ represents the correction to the asymptotic expectation value (which is $1+M$ for $w_{\mu \boldsymbol{\theta}}$). Equivalently, the Bartlett-corrected profile likelihood ratio is
\begin{equation}
\label{eq:pwstarDef}
    w_{\mu}^\ast = \frac{w_{\mu}}{\text{E}[w_{\mu}]} \equiv  \frac{w_{\mu}}{1 + b_{\mu}}\,,
\end{equation}
where $b_{\mu} = \text{E}[w_{\mu}] - 1$. The Bartlett-corrected $w_{\mu \boldsymbol{\theta}}^\ast$ and $w_{\mu}^\ast$ follow $\chi^2$ distributions with reduced errors of order $\varepsilon_s^4$ (see~\cite{bib:Canonero2023}).

One can use $w_{\mu}^\ast$ to compute confidence intervals for $\mu$ by imposing \(w_{\mu}^\ast < 1\) or, equivalently, \( \ell(\hat{\mu}, \hat{\boldsymbol{\theta}}) - \ell(\mu, \hat{\hat{\boldsymbol{\theta}}}) < \frac{1}{2} (1 + b_{\mu}) \).

In the GVM likelihoods considered here, $\mu$ and $\boldsymbol{\theta}$ are location parameters, meaning that $b_{\mu \boldsymbol{\theta}}$ and $b_{\mu}$ are constants because the values of $w_{\mu \boldsymbol{\theta}}$ and $w_{\mu}$ do not depend on $\mu$ or $\boldsymbol{\theta}$. Therefore, their expectation values are computed at the MLE values only.

The correction factors \( b_{\mu \boldsymbol{\theta}} \) and \( b_{\mu} \) can be estimated analytically to order \( \varepsilon_s^2 \). This perturbative calculation is carried out using the Lawley formula~\cite{bib:Lawley1956, bib:Codeiro1993}. For the likelihood defined in Eq.~\eqref{eq:gvm_avareges}, one can show that
\begin{equation}
\label{eq:b_GVM1}
    b_\mu = b_{\mu\boldsymbol{\theta}} - \Tilde{b}_{\boldsymbol{\theta}}\,,
\end{equation}
following the methodology detailed in~\cite{bib:Canonero2023}. The factors \( b_{\mu\boldsymbol{\theta}} \) and \( \Tilde{b}_{\boldsymbol{\theta}} \) are found to be
\begin{equation}
\label{eq:b_factors1}
\begin{split}
    b_{\mu\boldsymbol{\theta}} &= \sum_{s=1}^M \left[ 4\frac{j^{\theta_s \theta_s}}{\widehat{\sigma_{u_s}^2}} - \left(\frac{j^{\theta_s \theta_s}}{\widehat{\sigma_{u_s}^2}}\right)^2 \right] \varepsilon_s^2\,, \\
    \Tilde{b}_{\boldsymbol{\theta}} &= \sum_{s=1}^M \left[ 4\frac{\Tilde{j}^{\theta_s \theta_s}}{\widehat{\sigma_{u_s}^2}} - \left( \frac{\Tilde{j}^{\theta_s \theta_s}}{\widehat{\sigma_{u_s}^2} }\right)^2 \right] \varepsilon_s^2\,.
\end{split}
\end{equation}
Here, \( \widehat{\sigma_{u_s}^2} \) denotes the MLE of \( \sigma_{u_s}^2 \), obtained by evaluating Eq.~\eqref{eq:sigma_mle1} at \( \hat{\theta}_s \). The matrix \( j^{-1} \) represents the covariance matrix of the MLEs derived from the original quadratic likelihood in Eq.~\eqref{eq:comb_log_lik}, or equivalently Eq.~\eqref{eq:gvm_avareges} in the limit \(\varepsilon_s \rightarrow 0\). It is computed as:
\begin{equation}
\label{eq:cov_matrix_approach1}
j^{-1} = \begin{pmatrix}
j^{\mu \mu} & j^{\mu \boldsymbol{\theta}} \\[0.2cm]
j^{\boldsymbol{\theta} \mu} & j^{\boldsymbol{\theta} \boldsymbol{\theta}}
\end{pmatrix} = - \left( \nabla^2_{\mu,\, \boldsymbol{\theta}}\, \ell \bigg|_{\hat{\mu},\, \hat{\boldsymbol{\theta}}} \right)^{-1},
\end{equation}
where \( \nabla^2_{\mu,\, \boldsymbol{\theta}}\, \ell \) denotes the matrix of second derivatives of the log-likelihood with respect to all parameters \( (\mu, \boldsymbol{\theta}) \). Upper indices are used to label the components of \( j^{-1} \). The matrix \(\Tilde{j}^{-1}\) is defined by treating \( \mu \) as fixed,
\begin{equation}
\label{eq:cov_matrix_approach2}
\Tilde{j} = - \left(\nabla^2_{ \boldsymbol{\theta}}\, \ell\bigg|_{\, \hat{\boldsymbol{\theta}}}\right)^{-1}\,,
\end{equation}
and it is also computed using the quadratic likelihood given in Eq.~\eqref{eq:comb_log_lik}, or equivalently Eq.~\eqref{eq:gvm_avareges} in the limit \(\varepsilon_s \rightarrow 0\). \(j\) and \(\Tilde{j}\) depend on \(\sigma_{u_s}^2\) and are evaluated at \(\widehat{\sigma_{u_s}^2}\).
 
Such factors are generally small but can become non-negligible for values of \(\varepsilon_s\) exceeding about \(0.2\) or \(0.3\), and they converge to zero as \(\varepsilon_s \rightarrow 0\). They also remain small when \(b_{\mu\boldsymbol{\theta}} \sim \Tilde{b}_{\boldsymbol{\theta}}\), which happens when the MLEs of \(\mu\) and \(\boldsymbol{\theta}\) are uncorrelated. Further details on applying the Lawley formula to the GVM are given in~\cite{bib:Canonero2023}, where the Bartlett correction factor \(b_\mu\) was found to be zero under the assumption that the MLEs of \(\mu\) and \(\boldsymbol{\theta}\) are uncorrelated. Equation~\eqref{eq:b_GVM1} refines this result by providing a more precise correction.

These \( \varepsilon_s^2 \) approximations are highly effective when the condition specified in Eq.~\eqref{eq:condition1} is satisfied.

\subsection{Goodness-of-fit}
\label{sec:GOF}
To evaluate how well the selected model fits the observed data, a Goodness-of-Fit (GOF) statistic can be used. This can be defined as
\begin{equation}
    \label{eq:simpleAverage_GoF}
    q = -2\log\frac{L(\hat{\mu}, \hat{\boldsymbol{\theta}})}{L_{\rm s}(\hat{\boldsymbol{\varphi}}, \hat{\boldsymbol{\theta}})}\,,
\end{equation}
where \( L_{\rm s} \) represents the likelihood of the \textit{saturated model}. In the saturated model, the expected values \( \text{E}[y_i] = \mu \) are replaced by independent parameters \( \boldsymbol{\varphi} = (\varphi_1, \dots, \varphi_N) \), such that \( \text{E}[y_i] = \varphi_i \). Since there is one free parameter \( \varphi_i \) for each observed value \( y_i \), the log-likelihood of the saturated model is zero for Eq.~\eqref{eq:gvm_avareges}. Therefore, the GOF statistic simplifies to
\begin{equation}
\label{eq:q1}
\begin{split}
q=&\sum_{i,j=1}^N\left(y_i-\hat{\mu}-\sum_{s=1}^M\Gamma_i^{s}\hat{\theta}_s\right) V_{ij}^{-1} \left(y_j-\hat{\mu}-\sum_{s=1}^M\Gamma_j^{s}\hat{\theta}_s\right)\\& + \sum_{s=1}^M\left(1+\frac{1}{2\varepsilon_s^2}\right)\log\left[1+2\varepsilon_s^2\frac{(u_s-\hat{\theta}_s)^2}{v_s}\right]\,.
\end{split}
\end{equation}
To assess how well the model fits the observed data, one can compute a $p$-value from the GOF statistic. In the limit as \(\varepsilon_s \to 0\), the GOF statistic reduces to a sum of squares and follows a \(\chi^2\) pdf with \(N-1\) degrees of freedom. The $p$-value can then be calculated as \( p =  1 - F_{\chi^2}(q_{\rm obs}) \), where \(F_{\chi^2}\) is the cumulative distribution function of a \(\chi^2\) distribution, and \(q_{\rm obs}\) is the value of $q$ computed with the observed data.  Another common metric to assess the quality of the fit is the ratio \( q/N_{\text{dof}} \). If \( q \) follows a chi-squared distribution with \( N_{\text{dof}} \) degrees of freedom, its expected value is \( N_{\text{dof}} \), with a standard deviation of \( \sqrt{2N_{\text{dof}}} \). Therefore, if the model reasonably describes the data, this ratio is expected to be close to $1$.

For \(\varepsilon_s > 0\), the \(\chi^2\) distribution approximates the true distribution, with an error of order \(\varepsilon_s^2\), similar to the case of the likelihood ratio. To improve this approximation, we can apply a Bartlett correction to the GOF statistic, defined as
\begin{equation}
\label{eq:qstar}
    q^{\ast} = q \, \frac{N-1}{E[q]}= \frac{q}{1 + b_q/(N-1)},
\end{equation}
where \(E[q]\) is the expected value of \(q\) and $b_q = E[q] - N + 1$. In the limit as \(\varepsilon_s \to 0\), \(b_q \to 0\), and thus \(q^{\ast}\) converges to \(q\). For values of \(\varepsilon_s > 0\), the corrected statistic \(q^{\ast}\) follows a \(\chi^2\) distribution with \(N-1\) degrees of freedom, up to error terms of order \(\varepsilon_s^4\). Therefore, the corrected $p$-value 
$p^{\ast} =  1 - F_{\chi^2}(q^{\ast}_{\rm obs})$ provides a more reliable measure of how well the model fits the data. Alternatively, one can use the corrected ratio \( q^{\ast}/N_{\text{dof}} \) to assess the quality of the fit.

The factor \( b_q \) can be computed analytically to order \( \varepsilon_s^2 \) by expressing the expectation value of \( q \) as \( \text{E}[q] = -2\text{E}[\ell(\mu, \boldsymbol{\theta})] - \text{E}[w_{\mu\boldsymbol{\theta}}]\). The term $2\text{E}[\ell(\mu, \boldsymbol{\theta})]$ is computed using equation~$(3)$ from~\cite{bib:Canonero2023}. For the model of Eq.~\eqref{eq:gvm_avareges}, the result is
\begin{equation}
    b_q = 3\sum_{s=1}^M \varepsilon_s^2 - b_{\mu\boldsymbol{\theta}}\,,
\end{equation}
where \( b_{\mu\boldsymbol{\theta}} \) is defined in Eq.~\eqref{eq:b_factors1}.

\section{Application to $7$-$8$ TeV ATLAS-CMS top quark mass combination}
\label{sec:top_mass}

\begin{table*}[t]
\centering
\begin{tabular}{|c|c|c|c|}
\hline
$t\bar{t}$ final state & CM Energy $(\text{TeV})$ & $m_{top}\pm(\text{stat})\pm(\text{syst})\,(\text{GeV})$ & Total uncertainty $(\text{GeV})$ \\
\hline
All-hadronic~\cite{bib:ATLAS2014} & 7 & $175.06\pm 1.35\pm 1.21$ & $\pm 1.82$ \\
Dileptonic~\cite{bib:ATLAS2015} & 7 & $173.79\pm 0.54\pm 1.31$ & $\pm 1.42$ \\
Lepton+jets~\cite{bib:ATLAS2015} & 7 & $172.33\pm 0.75\pm 1.04$ & $\pm 1.28$ \\
All-hadronic\cite{bib:ATLAS2017} & 8 & $173.72\pm 0.55\pm 1.02$ & $\pm 1.16$ \\
Dileptonic~\cite{bib:ATLAS2016} & 8 & $172.99\pm 0.41\pm 0.74$ & $\pm 0.84$ \\
Lepton+jets~\cite{bib:ATLAS2018fwq} & 8 & $172.08\pm 0.39\pm 0.82$ & $\pm 0.91$ \\
\hline
\end{tabular}
\caption{Top quark mass measurements in $t\bar{t}$ final states at different center-of-mass energies, based on ATLAS data}
\label{tab:topmass_1}
\end{table*}

\begin{table*}[t]
\centering
\begin{tabular}{|c|c|c|c|}
\hline
$t\bar{t}$ final state & CM Energy $(\text{TeV})$ & $m_{top}\pm(\text{stat})\pm(\text{syst})\,(\text{GeV})$ & Total uncertainty $(\text{GeV})$ \\
\hline
All-hadronic~\cite{bib:CMS2014} & 7 & $173.49\pm 0.69\pm 1.23$ & $\pm 1.41$ \\
Dileptonic~\cite{bib:CMS2012a} & 7 & $172.50\pm 0.43\pm 1.52$ & $\pm 1.58$ \\
Lepton+jets~\cite{bib:CMS2012b} & 7 & $173.49\pm 0.43\pm 0.97$ & $\pm 1.06$ \\
All-hadronic\cite{bib:CMS2016a} & 8 & $172.32\pm 0.25\pm 0.57$ & $\pm 0.62$ \\
Dileptonic~\cite{bib:CMS2016a} & 8 & $172.22\pm 0.18\pm 0.94$ & $\pm 0.95$ \\
Lepton+jets~\cite{bib:CMS2016a} & 8 & $172.35\pm 0.16\pm 0.45$ & $\pm 0.48$ \\
Single top~\cite{bib:CMS2017} & 8 & $172.95\pm 0.77\pm 0.93$ & $\pm 1.20$ \\
$J/\psi$~\cite{bib:CMS2016b} & 8 & $173.50\pm 3.00\pm 0.94$ & $\pm 3.14$ \\
Secondary vertex~\cite{bib:CMS2016c} & 8 & $173.68\pm 0.20\pm 1.11$ & $\pm 1.12$ \\
\hline
\end{tabular}
\caption{Top-quark mass measurements in $t\bar{t}$ final states at different center-of-mass energies, based on CMS data}
\label{tab:topmass_2}
\end{table*}

\begin{figure*}[t]
    \begin{center}
    \includegraphics[width=0.9\textwidth]{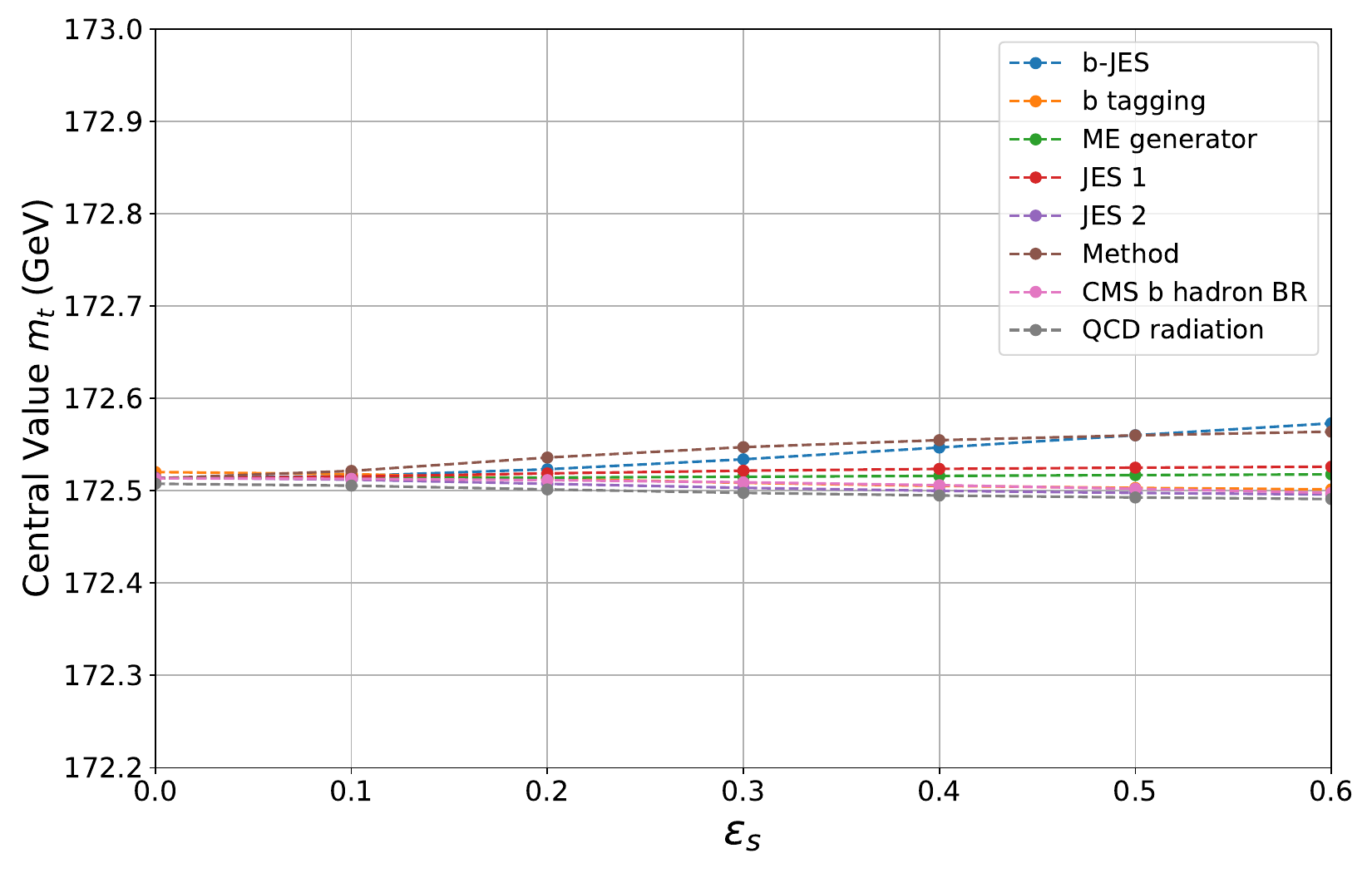}

    \caption{The plot shows the variation of the central value as a function of the \textit{error-on-error} parameter $\varepsilon_s$. Each line represents the change of the central value when the systematic uncertainties in the legend are considered uncertain one at a time. The central values are computed explicitly at points marked by dots and linearly interpolated in between}
    \label{fig:CV_original}
    \end{center}
\end{figure*}

\begin{figure*}[t]
    \begin{center}
    \includegraphics[width=0.9\textwidth]{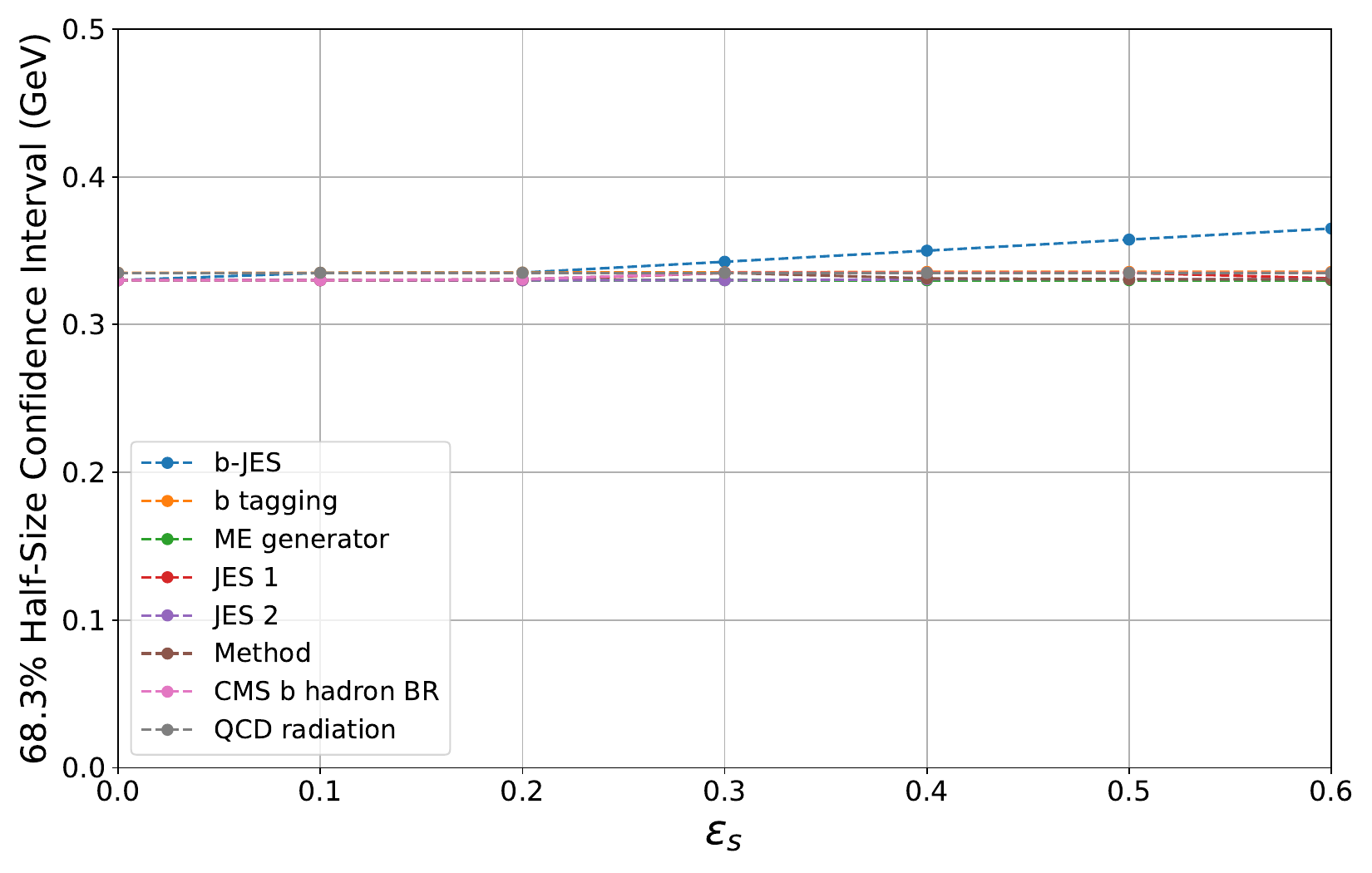}

    \caption{The plot shows the variation of the $68.3\%$ half-size confidence interval as a function of the \textit{error-on-error} parameter $\varepsilon_s$. Each line represents the change of the confidence interval when the systematic uncertainties in the legend are considered uncertain one at a time. The confidence intervals are computed explicitly at points marked by dots and linearly interpolated in between}
    \label{fig:CI_original}
    \end{center}
\end{figure*}

Measurements and combinations of the top-quark mass are a relevant testing ground for the \textit{errors-on-errors} framework for several reasons. Firstly, these measurements are becoming increasingly dominated by systematic uncertainties, making a precise evaluation of them crucial for the correct interpretation of results. Furthermore, measurements of the top-quark mass are often significantly impacted by QCD modeling uncertainties that themselves can be quite uncertain, especially those related to jets in the final state. In addition, the GVM's results are sensitive to the internal compatibility of the data, and some of the top-mass measurements exhibit tensions between each-other; for example, the Tevatron~\cite{bib:Tevatron2014} combination is slightly incompatible with the result of ATLAS-CMS one \cite{bib:ATLAS/CMS}, as well as the run $2$ ATLAS top-mass measurement exploiting a leptonic invariant mass~\cite{bib:ATLAS2023}.

We begin by replicating the results of the 7-8 TeV ATLAS-CMS top-mass combination using the BLUE\\ method, as described in Sec.~\ref{sec:BLUE}. The ATLAS inputs to the combination~\cite{bib:ATLAS2018fwq, bib:ATLAS2015, bib:ATLAS2016, bib:ATLAS2014, bib:ATLAS2017} are summarized in Tab.~\ref{tab:topmass_1}, while the CMS ones~\cite{bib:CMS2014, bib:CMS2012a, bib:CMS2012b, bib:CMS2016a, bib:CMS2017, bib:CMS2016b, bib:CMS2016c} are listed in Tab.~\ref{tab:topmass_2}.

To construct the BLUE covariance matrix, we use\\ Eqs.~\eqref{eq:V_blue} and~\eqref{eq:blue_cov_extened}, incorporating data from tables A.2 and A.3 in the appendices of~\cite{bib:ATLAS/CMS} (Arxiv version) and appendix B of~\cite{bib:ATLAS/CMSb}. Specifically, the statistical covariance matrix, which is diagonal, incorporates the reported statistical uncertainties for each measurement. Additionally, each term $\Gamma^s_i$ corresponds to the error on the $i$-th measurement attributed to systematic source $s$. As discussed previously, we redefine the parameters such that $\sigma^2_{u_s}=1$. The correlation coefficients $\rho^{(s)}_{ij}$ are determined using the correlation matrices from appendix~B of~\cite{bib:ATLAS/CMSb}. The result found using the BLUE approach is
\begin{equation}
    m_t = 172.51\pm 0.33 \, \text{GeV} \,.
\end{equation}
Our result has a minor discrepancy of $0.01$ GeV from the central value reported in \cite{bib:ATLAS/CMS}, but the confidence interval agrees with that reported in the paper. Discrepancies of this magnitude can be expected since all public results are rounded to the second decimal place.

\subsection{GVM analysis for combined estimate of $m_t$}
\label{sec:mtapp}
The objective of this section is to extend the top-mass combination to account for \textit{errors-on-errors}. This extension aims to evaluate the robustness of the combination against potential uncertainties in the assignment of systematic errors. We approach this in the simplest way possible, utilizing what we consider a suitable initial application of the GVM in situations where it is not obvious how to assign \textit{errors-on-errors}, or even whether it is necessary to do so. Specifically, we analyze each major systematic uncertainty individually, assuming that only one systematic is associated with an \textit{error-on-error} at a time. For each selected systematic, we vary the corresponding \textit{error-on-error} parameter to evaluate how this influences the results of the analysis. This process provides a clear understanding of the sensitivity of the combination to uncertainties in the assignment of systematic errors.

In this study, we consider as potentially uncertain the eight largest systematics as listed in Table~2 of~\cite{bib:ATLAS/CMS}, as we expect that the largest systematic errors would have a more pronounced effect on the final result if they are themselves uncertain. This selection does not imply that we consider these systematics as necessarily uncertain; rather, we intend to demonstrate the potential impact of various assumptions on the results of the combination.

To incorporate GVM-based systematic uncertainties, we first isolate the chosen systematic source \(s\) that we want to treat as uncertain and apply the NP approach as described in Sec.~\ref{sec:GVMcorr}, but with a few differences. We only use NPs to treat the systematic uncertainty with an associated \textit{error-on-error}, while all the other systematic uncertainties are incorporated in a BLUE-like covariance matrix. Furthermore, we introduce one NP for each set of non-trivial correlations (i.e., those not equal to \(\pm 1\)) induced across the measurements. If one systematic induces a correlation of \(\pm 1\) between two measurements, we use a single NP to model its effect in these two measurements. The resulting profile log-likelihood is given by
\begin{equation}
\label{eq:top_comb_log_lik}
\begin{split}
     \ell_p(\mu, \boldsymbol{\theta}_s)
    &= -\frac{1}{2} \sum_{i,j=1}^N 
    \bigl[ 
       y_i 
       - 
       \mu
       - 
       \sum_{e=1}^{M_s} \Gamma_{i,e}^{s} \,\theta_s^e 
    \bigr]\,
    \widetilde{W}_{ij}^{-1}\\ &\times \,
    \bigl[ 
       y_j 
       - 
       \mu
       - 
       \sum_{e=1}^{M_s} \Gamma_{j,e}^{s} \,\theta_s^e 
    \bigr]
    \\
    & - \frac{1}{2} \left( M_s + \frac{1}{2\varepsilon_s^2} \right)\log \left[
    1 + \frac{2\varepsilon_s^2}{v_s}\sum_{e,f=1}^{M_s}
    (\theta_s^e - u_s^e) \right.
    \\ & \times \left. 
    \bigl(\rho^{(s)}\bigr)^{-1}_{ef}\,
    (\theta_s^f - u_s^f)\right]\,.
\end{split}
\end{equation}
Here, \(M_s\) is the number of NPs introduced for source \(s\) and the term \(\Gamma_{i,e}^{s}\) represents the systematic shift in measurement \(i\) induced by the \(e\)-th NP. We set \(u_s^e = 0\) and \(v_s = 1\), which fixes the factors \(\Gamma_{i,e}^{s}\) to the systematic uncertainty values reported in Tables~A.2 and A.3 of~\cite{bib:ATLAS/CMS} (ArXiv version). The matrix \(\rho^{(s)}\) is taken from Appendix~B of~\cite{bib:ATLAS/CMSb}; before using it, we remove any dimensions corresponding to off-diagonal entries of \(\pm 1\), thereby reducing the dimensionality of the NP basis. 

To better understand how we introduce NPs, consider the largest systematic source in the combination: the \(b\)-JES uncertainty. This systematic arises from modeling the jet energy scale for \(b\)-jets in ATLAS and CMS. Within each experiment, the \(b\)-JES is either fully correlated or anti-correlated across its own measurements; however, it is only partially correlated between experiments, with a correlation coefficient of \(0.85\). We treat the set of ATLAS measurements as a single group (labeled \(e=1\)) and the set of CMS measurements as another group (labeled \(e=2\)). In this way, \(M_s = 2\), meaning we introduce two NPs: one for the ATLAS group and one for the CMS group. The factors \(\Gamma_{i,e}^{s}\) then quantify how each group’s NP shifts measurement \(i\). If a \(y_i\) is in the ATLAS group, \(\Gamma_{i,1}^{s}\neq0\) while \(\Gamma_{i,2}^{s}=0\), and vice versa for the CMS group. Consequently, each measurement can only belong to one group at a time. The sign of the factors \(\Gamma_{i,e}^{s}\) is determined by the sign of the \(\pm 1\) correlation within each group. Finally, the correlation matrix \(\rho^{(s)}\) is of dimension \(2\), with
\(\rho^{(s)}_{11} = \rho^{(s)}_{22} = 1\) and
\(\rho^{(s)}_{12} = \rho^{(s)}_{21} = 0.85\).

Some of the correlation matrices have negative eigenvalues, which we correct by adding the absolute value of the smallest negative eigenvalue to their diagonals. Although this has a minimal impact on the final combination, it is crucial for ensuring the convergence of the log-likelihood maximization.

The final term we must specify in Eq.~\eqref{eq:top_comb_log_lik} is the BLUE-like covariance matrix
\[
    \widetilde{W}_{ij} 
    \;=\;
    V_{ij} 
    \;+\; 
    \sum_{\substack{p=1 \\ p \neq s}}^M U_{ij}^{(p)}\,.
\]
We build this covariance matrix as described at the beginning of this section, where we reproduced the combination result using the BLUE approach. The only difference is that we now remove the contribution of source \(s\) from the covariance matrix, since it is handled through NPs instead.

The log-likelihood is maximized with respect to the NPs using the perturbative approach described in Sec.~\ref{sec:central_values}, while employing the formulas provided in App.~\ref{app:B} to account for non-trivial correlations, adapted to the structure of Eq.~\eqref{eq:top_comb_log_lik}.

\subsection{Results}
Figure~\ref{fig:CV_original} illustrates the variation in the central value (i.e., Maximum Likelihood Estimator) of the combination when one of the systematic uncertainties listed in the legend is itself considered uncertain. Specifically, the plot shows the dependence of the central value \(\hat{\mu}\) (here $\mu = m_t$) on the \textit{error-on-error} parameter \(\varepsilon_s\) for each systematic uncertainty. The nomenclature for the systematic components utilized here is the same used in \cite{bib:ATLAS/CMS}, where a detailed explanation of what they describe can be found. The most important conclusion here is that the central value of the combination is quite robust to the presence of uncertain systematic errors. Specifically, the change in the central value remains always within $0.1$ GeV for the explored range of \(\varepsilon_s\) values, which is well within the confidence interval of approximately $0.3$ GeV. Ultimately, all central values converge to the BLUE result as \(\varepsilon_s\) approaches zero, meaning that the way we implement correlations in our model, as described in Sec.~\ref{sec:BLUE}, is consistent. The reason why the \textit{b tagging} and \textit{QCD radiation} lines do not converge to the same point is due to the regularization of their associated correlation matrices, as previously described.

Similarly, Fig.~\ref{fig:CI_original} displays the half-width of the $68.3\%$ confidence interval as a function of $\varepsilon_s$. As for the central value, each line corresponds to the case where only one systematic uncertainty at a time is considered as itself uncertain. The confidence intervals were computed using the Bartlett corrections, as described in Sec.~\ref{sec:CI}, while employing the formulas provided in App.~\ref{app:B} to account for non-trivial correlations.

Figure~\ref{fig:CI_original} confirms our previous conclusion: the combination is generally robust to the presence of uncertain systematic errors. However, an increase of about $10\%$ in the confidence interval is observed when the \textit{b-JES} systematic uncertainty has a relative error of approximately $50\%$. The combination's heightened sensitivity to uncertainties in this systematic error stems from the fact that it constitutes the largest uncertainty of the combination. Notice that the \textit{b-JES} systematic uncertainty, modeling the flavor response of b-jets, is a two-point systematic, i.e., based on the difference obtained with two methods, and thus falls into the category of systematic sources that could plausibly be uncertain. The confidence interval remains stable if any other of the systematics considered are taken as themselves uncertain. A negligible bias in the size of the confidence interval stemming from the \textit{b tagging} and \textit{QCD radiation} uncertainties is present due to the regularization of the correspondent correlation matrices. 

\subsection{Sensitivity to outliers}
\label{sec:outliers}

\begin{figure*}[t!]
    \begin{center}
    \includegraphics[width=0.9\textwidth]{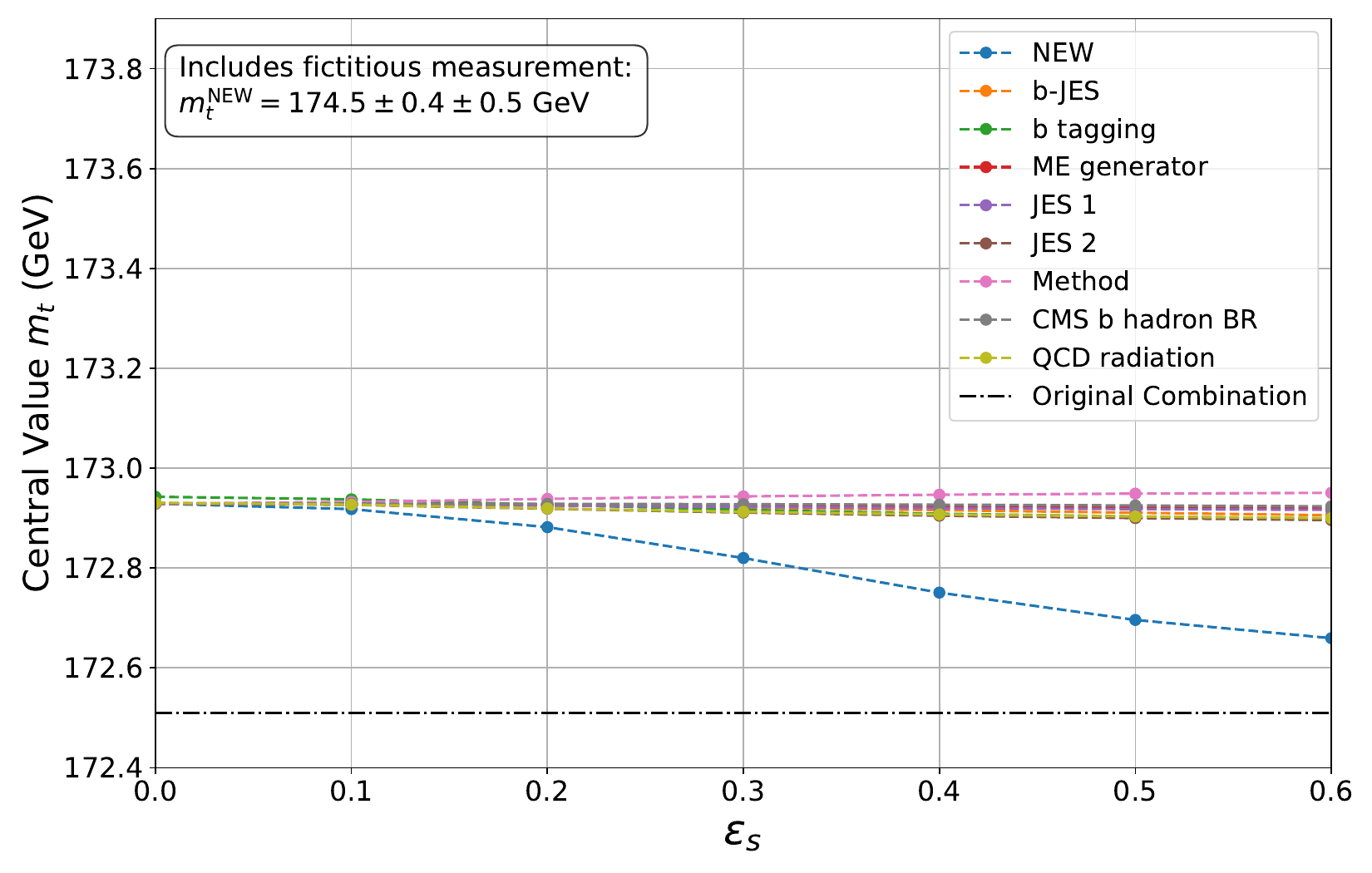}
    \caption{The plot shows the variation of the central value as a function of the \textit{error-on-error} parameter $\varepsilon_s$ when one fictitious measurement is included in the combination. Each line represents the change of the central value when the systematic uncertainties in the legend are considered uncertain one at a time. The central values are computed explicitly at points marked by dots and linearly interpolated in between}
    \label{fig:CV_new}
    \end{center}
\end{figure*}

\begin{figure*}[t!]
    \begin{center}
    \includegraphics[width=0.9\textwidth]{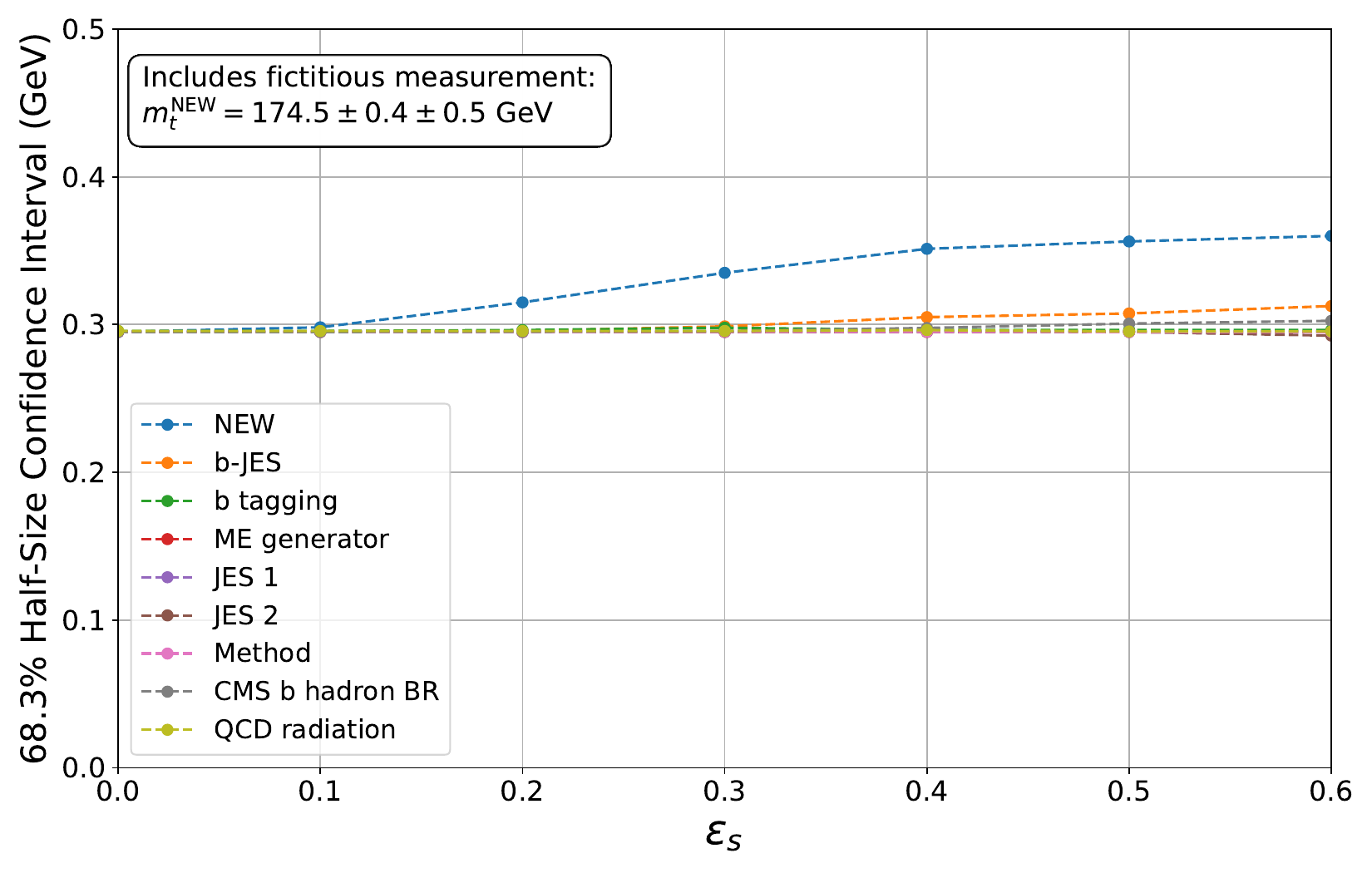}
    \caption{The plot shows the variation of the $68.3\%$ half-size confidence interval as a function of the \textit{error-on-error} parameter $\varepsilon_s$ when one fictitious measurement is included in the combination. Each line represents the change in the confidence interval when the systematic uncertainties in the legend are considered uncertain one at a time. The confidence intervals are computed explicitly at points marked by dots and linearly interpolated in between}
    \label{fig:CI_new}
    \end{center}
\end{figure*}

The conclusions of the last section would be significantly impacted if any of the combination inputs exhibited tension with the rest of the measurements. It is not uncommon to have input values that exhibit a significant tension and it could possibly happen for an updated LHC-Tevatron combination, or for a LHC Run $2$ combination that includes the ATLAS top-mass measurement exploiting a leptonic invariant mass~\cite{bib:ATLAS2023}. However, since extending the combination to include additional measurements would require knowledge of the correlations between them and those listed in Tabs.~\ref{tab:topmass_1} and~\ref{tab:topmass_2}, we introduce a fictitious measurement into the dataset to illustrate the properties of the GVM in such a scenario. Specifically, we consider adding a measurement with a central value of \(m_t^{\mathrm{NEW}} = 174.5\) GeV, a statistical uncertainty of \(0.4\) GeV, and a global systematic uncertainty of \(0.5\) GeV. We assume that the systematic uncertainty of the new measurement is uncorrelated with the uncertainties of the other measurements.

If the fictitious measurement is added to the combination, without considering \textit{errors-on-errors}, the result obtained using the BLUE approach is \( m_t = 172.91 \pm 0.29 \, \text{GeV} \). As anticipated, the inclusion of the new measurement shifts the central value of the combination towards higher values and results in a reduction of the confidence interval.

Figure~\ref{fig:CV_new} illustrates the variations in the central value of the combination when one of the systematic uncertainties listed in the legend is itself considered uncertain. On top of the systematics considered in the last section, the \textit{NEW} systematic uncertainty of the fictitious measurement is also treated as potentially uncertain. In contrast to observations made in the previous section, the inclusion of the outlier in the dataset makes the effect of \textit{errors-on-errors} non-negligible. The central value significantly shifts when the \textit{NEW} systematic uncertainty is considered uncertain, moving back to that of the original combination. This is an important property of the GVM: while typically the presence of an outlier would significantly pull the result of a combination, within the GVM framework as the \textit{error-on-error} parameter increases, the outlier is assigned lesser weight within the combination. Consequently, the central value shifts back towards that of the original combination, thereby diminishing the outlier's effect.

Similarly, Fig.~\ref{fig:CI_new} displays the variation in the half-size of the \(68.3\%\) confidence interval as a function of \(\varepsilon_s\). The plot shows how the inclusion of the fictitious measurement affects the combination. It demonstrates that when the \textit{NEW} systematic uncertainty includes an associated \textit{error-on-error}, the confidence interval can increase significantly, by up to 25\%. This is the second relevant property of the GVM: The size of the confidence interval is sensitive to the internal consistency of the input data. Specifically, the less compatible the input measurements are, the more the confidence interval inflates when \textit{errors-on-errors} are considered. This is because the GVM treats the tension in the dataset as an additional source of uncertainty resulting in an increase in the size of the confidence interval.

\begin{figure*}[t!]
\begin{center}
\includegraphics[width=0.49\textwidth]{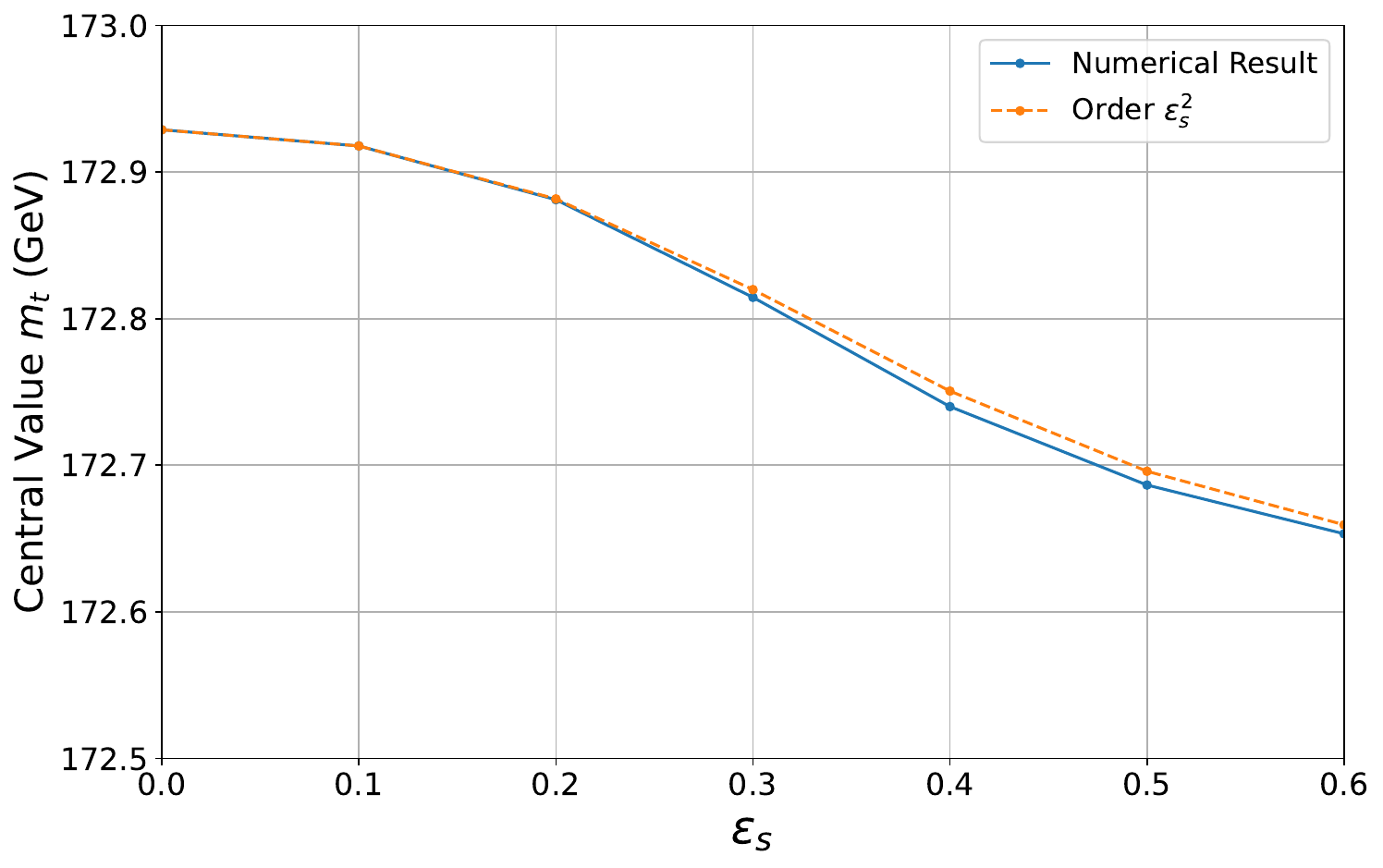}
\includegraphics[width=0.49\textwidth]{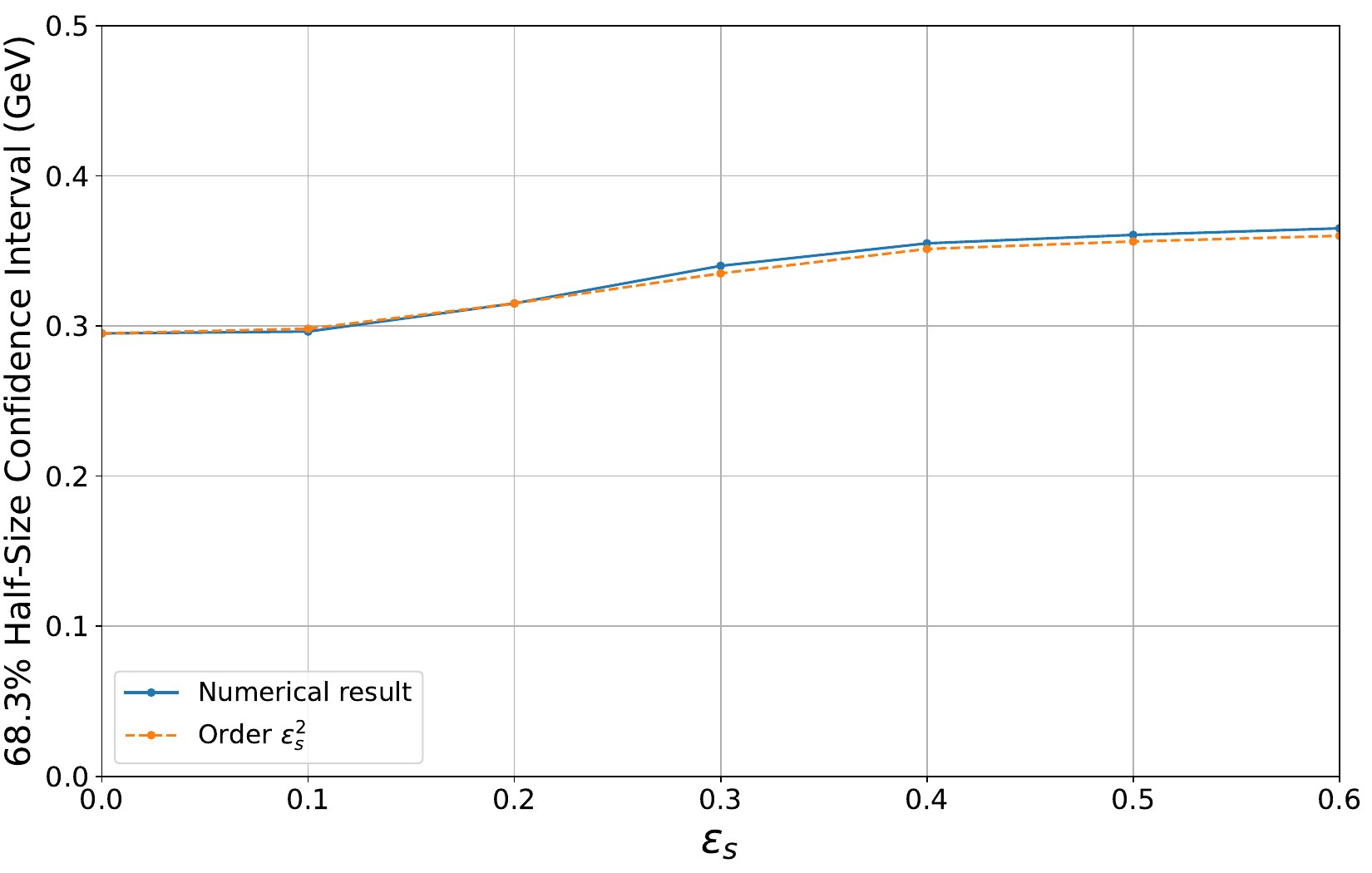}
\caption{Comparison of analytical perturbative results at order \(\varepsilon_s^2\) (orange line) and numerical results (blue line)}
\label{fig:comparison}
\end{center}
\end{figure*}

The perturbative analytical methods used to derive our results proved to be highly precise, even when the \textit{NEW} systematic uncertainty was treated as uncertain and the conditions for applying perturbative techniques, as outlined in Eq.~\eqref{eq:condition2}, were not fully met. This is illustrated in Fig.~\ref{fig:comparison}, where we compare the analytical results obtained using the perturbative approach outlined in Sec.~\ref{sec:useful-formulas} at orders \(\varepsilon_s^2\) with those from a numerical approach. This comparison focuses on the case where the \textit{NEW} systematic uncertainty is treated as uncertain. The numerical results were obtained by numerically maximizing the likelihood over all parameters and calculating the expectation value of the profile likelihood ratio to determine the Bartlett correction, using MC simulations. Details of the simulation process are provided in App.~\ref{app:C}.

\section{Conclusions}
\label{sec:conclusions}
The Gamma Variance Model (GVM) provides a powerful statistical framework for addressing uncertainties in the assignment of systematic errors, informally referred to as \textit{errors-on-errors}.  Going beyond previous publications \cite{bib:Cowan2019,bib:Canonero2023}, we have extended the original model to include non-trivial correlations (not only $0$ or $\pm 1$) and have provided formulas that simplify its application in practice by avoiding the need for numerical methods.  We also derived a useful connection between the BLUE method for combining results under non-trivial correlations and the corresponding likelihood method using nuisance parameters. This connection was employed here to implement \textit{errors-on-errors} when systematic uncertainties induce non-trivial correlations, but it can also be applied in other contexts.

We argue that the assumption of independent gamma distributions for estimates of variances is a plausible model that arises, e.g., if nuisance parameters are estimated through an average of Gaussian-distributed measurements.  In a more general scenario, this assumption does not necessarily hold but the model should nevertheless provide more realistic inference than if one were to treat assigned variances as exactly known.  The results one might obtain if the variance estimates followed some other bell-shaped distribution are expected to be qualitatively similar but dependent in general on the specific problem.  Thus the precision of the inferences derived from the GVM is limited, but the model remains useful for identifying trends and understanding how an analysis would shift when uncertainties in the assignment of systematic errors are substantial.

We applied the framework to the 7-8 TeV ATLAS-CMS top quark mass combination \cite{bib:ATLAS/CMS}. All the results have been studied by considering various systematic uncertainties as uncertain, one at a time, and varying their associated \textit{error-on-error} parameters $\varepsilon_s$. The aim of this paper has been to illustrate the impact of \textit{errors-on-errors} on the combination, rather than assigning precise uncertainties to the systematic errors. This methodology can be used as a general procedure to identify the systematic uncertainties to which a combination is sensitive when they have associated \textit{errors-on-errors}, and as a general approach to assess the robustness of a combination against systematic uncertainties that are themselves uncertain.

We conclude that the ATLAS-CMS $m_t$ combination is robust, with the central value remaining very stable across a broad spectrum of assumed values for \textit{errors-on-errors}. The confidence interval remains stable as well, though it exhibits up to a $10\%$ increase when the \textit{b-JES} systematic uncertainty is considered uncertain.

The scenario of an outlier present in the combination was also explored. This study is relevant to demonstrate the model's properties in situations that may occur in future combinations. This example demonstrated the model's sensitivity to the internal compatibility of the dataset. Specifically, the central value of the combination is less biased by outliers as the \textit{errors-on-errors} parameters increase. Meanwhile, the confidence interval inflates with rising internal inconsistency among the data, as the GVM treats internal inconsistency in the input data as an additional source of uncertainty.  


\section*{Acknowledgements}

The authors are grateful to Mark Owen, Véronique Bois\-vert, and colleagues in the ATLAS Collaboration for their valuable feedback on this work. We also thank Richard Lockhart and Sara Algeri for organizing the Phystat review on the GVM, as well as Matthew Reader and Robert Thorne for productive discussions on the \textit{errors-on-errors} framework. Finally, we acknowledge the support of the U.K.\ Science and Technology Facilities Council.

\section*{Data Availability Statement}

This manuscript utilizes data from previously published studies, specifically in Sec.~\ref{sec:top_mass}, where references are provided in the text. The remaining sections do not contain new data but are based on the results of calculations.

\appendix
\section{Profiling over NPs modeling correlated auxiliary measurements}
\label{app:A}
The BLUE log-likelihood of Eq.~\eqref{eq:blue}, with the covariance matrix defined as in Eq.\eqref{eq:blue_cov_extened}, can be derived by profiling Eq.\eqref{eq:comb_log_lik_u} over all the NPs $\theta_s^i$ when the auxiliary measurements $u_s^i$ are set to $0$. This process involves profiling the likelihood recursively over all $\theta_s^i$ for a given $s$. Here, we demonstrate how to do this for $s=M$. To simplify the notation we define
\begin{equation}
    \tilde{\mu}_i = \mu + \sum_{s=1}^{M-1}\Gamma_i^{s}\theta_s^i\,
\end{equation}
and 
\begin{equation}
    K = -\frac{1}{2}\sum_{s=1}^{M-1}\sum_{i,j=1}^N\frac{\theta^i_s\theta^j_s}{\sigma_{u_s}^2}\left(\rho^{(s)}\right)_{ij}^{-1}\,.
\end{equation}
With these choices Eq.\eqref{eq:comb_log_lik_u} becomes
\begin{equation}
\begin{split}
    \ell(\mu,\boldsymbol{\theta})&=-\frac{1}{2}\sum_{i,j=1}^N\left(y_i-\Tilde{\mu}_i-\Gamma_i^{M}\theta^i_M\right) V_{ij}^{-1} \\ &\times \left(y_j-\tilde{\mu}_j-\Gamma_j^{M}\theta^j_M\right)\\&-\frac{1}{2}\sum_{i,j=1}^N\frac{\theta^i_M\theta^j_M }{\sigma_{u_M}^2}\left(\rho^{(M)}\right)_{ij}^{-1}+ K \,.
\end{split}
\end{equation}
To further simplify the derivation we define $\Tilde{\theta}_i = \Gamma^M_i \theta^i_M$ and $\Tilde{\rho}^{(M)}_{ij} =  \rho^{(M)}_{ij}\Gamma^M_i\Gamma^M_j\sigma_{u_M}^2$. Thus the log-likelihood can be written in the simple form:
\begin{equation}
\begin{split}
    \ell(\mu,\boldsymbol{\theta})=&-\frac{1}{2}\sum_{i,j=1}^N\left(y_i-\Tilde{\mu}_i-\Tilde{\theta}_i\right) V_{ij}^{-1} \left(y_j-\tilde{\mu}_j-\Tilde{\theta}_j\right)\\&-\frac{1}{2}\sum_{i,j=1}^N\Tilde{\theta}_i\Tilde{\theta}_j \left(\Tilde{\rho}^{(M)}\right)_{ij}^{-1}+ K\,.
\end{split}
\end{equation}
To profile over the $\Tilde{\theta}_i$, we need to solve the following equations for $\Tilde{\theta}_i$ for all $i$:
\begin{equation}
    \frac{\partial \ell}{\partial \Tilde{\theta}_i} = \sum_j^N (y_j-\Tilde{\mu}_j) V_{ij}^{-1} - \Tilde{\theta}_j\left[ V_{ij}^{-1} + \left(\Tilde{\rho}^{(M)}\right)_{ij}^{-1}\right]= 0\,.
\end{equation}
This leads to the profiled values of the $\Tilde{\theta}_i$ parameters,
\begin{equation}
    \widehat{\widehat{\Tilde{\theta}_i}} = \sum_{j,k}^N M^{-1}_{ij}V^{-1}_{jk}(y_k-\Tilde{\mu}_k)\,,
\end{equation}
where the matrix $M$ is defined by 
\begin{equation}
    M_{ij} = V_{ij}^{-1} + \left(\Tilde{\rho}^{(M)}\right)_{ij}^{-1} \,.
\end{equation}
This matrix is symmetric and we will use this property to switch its indices for the proof. If we now substitute the profiled values $\hat{\hat{\Tilde{\theta}}}_i$ back into the log-likelihood we obtain 

\begin{equation}
\small
\begin{split}
    &\ell(\mu,\boldsymbol{\theta_1}, .., \boldsymbol{\theta_{M-1}}, \boldsymbol{\widehat{\widehat{\Tilde{\theta}}}_M}) = \\& -\frac{1}{2} \sum_{i,j=1}^N \left( y_i-\Tilde{\mu}_i \right)
    \left[ V_{ij}^{-1} - \sum_{p,q=1}^N 
    V_{ip}^{-1} M^{-1}_{pq}
    V_{qj}^{-1} \right]
    \left(y_j-\tilde{\mu}_j\right)+ K\,.
\end{split}
\end{equation}

\noindent To further simplify this expression, the Woodbury identity \cite{bib:woodbury} can be used to show that

\begin{equation}
\begin{split}
    \left[V^{-1}- 
    V^{-1}M^{-1}V^{-1}\right]_{ij}^{-1} &= 
V_{ij} - \left(-M + V^{-1}\right)_{ij}^{-1} 
    \\ &= V_{ij} + \Tilde{\rho}_{ij}^{(M)}\,.
\end{split}
\end{equation}

\noindent Therefore we find

\begin{equation}
\begin{split}
    \ell(\mu&,\boldsymbol{\theta_1}, .., \boldsymbol{\theta_{M-1}}, \boldsymbol{\widehat{\widehat{\Tilde{\theta}}}_M})=\\ & -\frac{1}{2}
    \sum_{i,j=1}^N \left(y_i-\Tilde{\mu}_i\right) 
    \left[ V + \Tilde{\rho}^{(M)} \right]_{ij}^{-1}
    \left( y_j - \tilde{\mu}_j \right)+ K\,,
\end{split}
\end{equation}

\noindent where $\Tilde{\rho}^{(M)}_{ij} = \rho^{(M)}_{ij}\Gamma^M_i\Gamma^M_j \sigma_{u_M}^2$ represents the $M$-th term in the systematic part of the BLUE covariance matrix $W$ as defined in Eq.~\eqref{eq:blue_cov_extened}, i.e., $U^{(M)} = \Tilde{\rho}^{(M)}$. This procedure can be recursively applied to compute all terms in the BLUE covariance matrix corresponding to every source of systematic uncertainty. Thus, one obtains $W = V + \sum_{s=1}^M \Tilde{\rho}^{(s)} = V + \sum_{s=1}^M U^{(s)}$.

\section{Useful formulas for the GVM with non-trivial correlations}
\label{app:B}
In this appendix, we provide useful formulas to facilitate the implementation of the GVM with non-trivial correlations, similar to the approach outlined in Sec.~\ref{sec:useful-formulas} for the case of trivial correlations.

\subsection{Profiled values of nuisance parameters}
The method outlined in Sec.~\ref{sec:central_values} for profiling the NPs can be extended to solve the score equations perturbatively for the GVM with arbitrary correlations, as defined in Eq.~\eqref{eq:prof_comb_log_lik_u}. Our aim is to solve
\begin{equation}
\small
\begin{split}
&\sum_{j=1}^N \Gamma_i^{s} V_{ij}^{-1} \left(y_j - \mu  - \sum_{p=1}^M \Gamma_j^{p} \hat{\hat{\theta}}_{f(p,j)} \right) \\ &+ \frac{(1 + 2N\varepsilon_s^2)\sum_{j=1}^N \left( \rho^{(s)}_{ij} \right)^{-1} \left( u_s^{j} - \hat{\hat{\theta}}_{f(s,j)} \right)}{v_s + 2\varepsilon_s^2 \sum_{i,j=1}^N \left( u_s^{i} - \hat{\hat{\theta}}_{f(s,i)} \right) \left( \rho^{(s)}_{ij} \right)^{-1} \left( u_s^{j} - \hat{\hat{\theta}}_{f(s,j)} \right)} = 0\,,
\end{split}
\end{equation}
where we introduce the composite index $f(s,i) = N(s-1) + i$ to denote $\theta_s^i$ for notational convenience. To solve the score equation, we define the perturbative series for $\hat{\hat{\theta}}_{f(s,i)}$ as
\begin{equation} 
\label{eq:expansion2}
\hat{\hat{\theta}}_{f(s,i)} = \hat{\hat{\theta}}_{f(s,i)}^{(0)} + \varepsilon_s^2 \hat{\hat{\theta}}_{f(s,i)}^{(1)} + \varepsilon_s^4 \hat{\hat{\theta}}_{f(s,i)}^{(2)} + \cdots \,.
\end{equation}
The first term in this series is given by
\begin{equation} 
\begin{split}
\hat{\hat{\theta}}_{f(s,i)}^{(0)} &= \sum_{p=1}^M\sum_{j=1}^N \left( C^{(0)} \right)^{-1}_{f(s,i)\, f(p,j)} \left[ \sum_{k=1}^N \Gamma_j^{p} V_{jk}^{-1} \left( y_k - \mu \right ) \right. \\ &\left.+ \sum_{k=1}^N \frac{\left( \rho^{(p)} \right)_{jk}^{-1} u_p^k}{v_p} \right]\,,
\end{split}
\end{equation}
where the matrix $C^{(0)}_{f(s,i)\, f(p,j)}$ is defined as
\begin{equation} 
C^{(0)}_{f(s,i)\, f(p,j)} = \Gamma_i^{s} V_{ij}^{-1} \Gamma_j^{p} + \frac{\delta_{sp}}{v_s} \left( \rho^{(s)} \right)_{ij}^{-1}\,.
\end{equation}
This first-order result is also the solution to the quadratic likelihood given by Eq.~\eqref{eq:comb_log_lik_u}. At a generic order $\varepsilon_s^{2n}$, with $n\geq 1$, the perturbative term $\varepsilon_s^{2n}\, \hat{\hat{\theta}}_{f(s,i)}^{(n)}$ is
\begin{equation}
\begin{split}
&\varepsilon_s^{2n}\hat{\hat{\theta}}_{f(s,i)}^{(n)} = \sum_{p=1}^M \sum_{j=1}^N\left( C^{(n)} \right)^{-1}_{f(s,i)\, f(p,j)}\\& \times\left[ \sum_{k=1}^N \Gamma_j^{p} V_{jk}^{-1} \left( y_k - \mu - \sum_{p'=1}^M \Gamma_k^{p'}\, T_{f(p',k),n-1} \right)\right.\\ &\quad + \left. \frac{ \sum_{k=1}^N \left( \rho^{(p)}\right)_{jk}^{-1} \left(u_p^k-T_{f(p,k),n-1}\right)}{S_{p,n-1}^{2}} \right] \,.
\end{split}
\end{equation}
Here, the matrix $C^{(n)}_{f(s,i)\, f(p,j)}$ is defined as
\begin{equation} 
C^{(n)}_{f(s,i)\, f(p,j)} = \Gamma_i^{s} V_{ij}^{-1} \Gamma_j^{p} + \frac{\delta_{sp}}{S_{s,n-1}^{2}} \left( \rho^{(s)} \right)_{ij}^{-1}\,,
\end{equation}
while \(T_{f(s,i),n} \) and \(S_{s,n}^{2}\) are defined as
\begin{equation}
    T_{f(s,i),n} = \hat{\hat{\theta}}_{f(s,i)}^{(0)} + \varepsilon_s^2 \hat{\hat{\theta}}_{f(s,i)}^{(1)} + \cdots + \varepsilon_s^{2n} \hat{\hat{\theta}}_{f(s,i)}^{(2n)}\,,
\end{equation}
\begin{equation}
\begin{split}
S_{s,n}^{2} &= \frac{1}{1 + 2N\varepsilon_s^2}\left[ v_s +2\varepsilon_s^2\sum_{i,j=1}^N \left( u_s^{i} - T_{f(s,i),n} \right)\right. \\ &\times \left( \rho^{(s)} \right)_{ij}^{-1}\left.\left( u_s^{j} -T_{f(s,j),n} \right)\right]\,.
\end{split}
\end{equation}
Notice that if the condition 
\begin{equation}
\label{eq:condition2}
\frac{2\varepsilon_s^2}{v_s} \sum_{i,j=1}^N \left( u_s^{i} - \hat{\hat{\theta}}_{f(s,i)} \right) \left( \rho^{(s)}_{ij} \right)^{-1} \left( u_s^{j} - \hat{\hat{\theta}}_{f(s,j)} \right) < 1
\end{equation}
is not satisfied, the perturbative series is not guaranteed to converge. 

When variations of this model are employed, the same perturbative procedure must be used. To solve the perturbative equation for \(\varepsilon_s^{2n} \hat{\hat{\theta}}_{f(s,i)}^{(n)}\), one substitutes \( \hat{\hat{\theta}}_{f(s,i)} \) with \(T_{f(s,i),n}\) throughout the score equation, except in the denominator where \(T_{f(s,i),n-1}\) is substituted instead (due to the presence of the \(\varepsilon_s^2\) multiplicative factor). Since all perturbative terms in \(T_{f(s,i),n-1}\) have already been determined recursively, one can isolate \(\varepsilon_s^{2n} \hat{\hat{\theta}}_{f(s,i)}^{(n)}\) and solve for it directly.

\subsection{Confidence intervals}
We can compute the Bartlett correction for the GVM generalized to handle any correlations using the same approach described in Sec.~\ref{sec:CI}. Specifically, for the likelihood defined by Eq.~\eqref{eq:prof_comb_log_lik_u}, the factor \( b_\mu \) can be expressed as
\begin{equation}
\label{eq:b_GVM2}
    b_\mu = b_{\mu\boldsymbol{\theta}} - \Tilde{b}_{\boldsymbol{\theta}}\,,
\end{equation}
where the terms \( b_{\mu\boldsymbol{\theta}} \) and \( \Tilde{b}_{\boldsymbol{\theta}} \) are defined as
\begin{equation}
\small
\label{eq:b_factors2}
\begin{split}
b_{\mu\boldsymbol{\theta}} &=  \sum_{s=1}^M \left\{ 4\frac{\text{Tr}\left[ j^{\boldsymbol{\theta}_s \boldsymbol{\theta}_s} \left( \rho^{(s)} \right)^{-1} \right]}{ \widehat{\sigma_{u_s}^2} } + \frac{ \left( \text{Tr}\left[ j^{\boldsymbol{\theta}_s \boldsymbol{\theta}_s} \left( \rho^{(s)} \right)^{-1} \right] \right)^2 }{ \left( \widehat{\sigma_{u_s}^2} \right)^2 }   \right. \\ &  \left. - 2\frac{\text{Tr}\left[ j^{\boldsymbol{\theta}_s \boldsymbol{\theta}_s} \left( \rho^{(s)} \right)^{-1} j^{\boldsymbol{\theta}_s \boldsymbol{\theta}_s} \left( \rho^{(s)} \right)^{-1} \right]}{ \left( \widehat{\sigma_{u_s}^2} \right)^2 }\right\} \varepsilon_s^2 \,,
\end{split}
\end{equation}
\begin{equation}
\small
\begin{split}
\label{eq:tilde_b_GVM2_modified}
    \Tilde{b}_{\boldsymbol{\theta}} &=  \sum_{s=1}^M \left\{ 4\frac{\text{Tr}\left[ \Tilde{j}^{\boldsymbol{\theta}_s \boldsymbol{\theta}_s} \left( \rho^{(s)} \right)^{-1} \right]}{ \widehat{\sigma_{u_s}^2} } + \frac{ \left( \text{Tr}\left[ \Tilde{j}^{\boldsymbol{\theta}_s \boldsymbol{\theta}_s} \left( \rho^{(s)} \right)^{-1} \right] \right)^2 }{ \left( \widehat{\sigma_{u_s}^2} \right)^2 } \right. \\ &  \left.- 2\frac{\text{Tr}\left[ \Tilde{j}^{\boldsymbol{\theta}_s \boldsymbol{\theta}_s} \left( \rho^{(s)} \right)^{-1} \Tilde{j}^{\boldsymbol{\theta}_s \boldsymbol{\theta}_s} \left( \rho^{(s)} \right)^{-1} \right]}{ \left( \widehat{\sigma_{u_s}^2} \right)^2 } \right\} \varepsilon_s^2 \,.
\end{split}
\end{equation}
Here, \( \widehat{\sigma_{u_s}^2} \) is the MLE of \( \sigma_{u_s}^2 \), which can be computed by evaluating Eq.~\eqref{eq:sigma_mle2} at \( \hat{\theta}_s^i \). The matrices \( j^{-1} \) and \( \Tilde{j}^{-1} \) are defined by Eqs.~\eqref{eq:cov_matrix_approach1} and~\eqref{eq:cov_matrix_approach2}, respectively, and are applied to the quadratic likelihood given in Eq.~\eqref{eq:comb_log_lik_u}. Additionally, \( j^{\boldsymbol{\theta}_s \boldsymbol{\theta}_s} \) and \( \Tilde{j}^{\boldsymbol{\theta}_s \boldsymbol{\theta}_s} \) are submatrices of \( j^{-1} \) and \( \Tilde{j}^{-1} \) that correspond to a fixed index \( s \). Both \( j^{-1} \) and \( \Tilde{j}^{-1} \) are functions of \( \sigma_{u_s}^2 \) and must be evaluated at \( \widehat{\sigma_{u_s}^2} \). To simplify the computation of the Bartlett correction, the double-indexed NP \( \theta_s^i \) can be flattened into single-indexed variable by defining \( f(s, i) = N(s-1) + i \), so that \( \theta_s^i = \theta_{f(s, i)} \). 

For variations of the log-likelihood definition, such as the one in Eq.~\eqref{eq:top_comb_log_lik}, the Bartlett factors retain the same functional form. One need only apply the matrices \(j^{-1}\) and \(\Tilde{j}^{-1}\) to the corresponding quadratic log-likelihood, which, for instance, arises from Eq.~\eqref{eq:top_comb_log_lik} in the limit \(\varepsilon_s \to 0\).

\subsection{Goodness-of-fit}
The GOF statistic of the GVM with non-trivial correlations (Eq.~\eqref{eq:prof_comb_log_lik_u}) is defined as
\begin{equation}
\label{eq:q2}
\begin{split}
    q&=\sum_{i,j=1}^N\left(y_i-\hat{\mu}-\sum_{s=1}^M\Gamma_i^{s}\hat{\theta}^i_s\right) V_{ij}^{-1} \left(y_j-\hat{\mu}-\sum_{s=1}^M\Gamma_j^{s}\hat{\theta}^j_s\right)\\&+\sum_{s=1}^M \left(N+\frac{1}{2\varepsilon^2_s}\right) \log \left[1 + \frac{2\varepsilon_s^2}{v_s} \sum_{i,j=1}^N(u^i_s-\hat{\theta}^i_s) \right.\\ & \times \left. \left(\rho^{(s)}\right)_{ij}^{-1}(u^j_s-\hat{\theta}^j_s)\right]\,.
\end{split}
\end{equation}
This GOF statistic can be corrected using the Bartlett correction as outlined in Sec.~\ref{sec:GOF}. Specifically the Bartlett factor is equal to
\begin{equation}
    b_q = (2N + N^2) \sum_{s=1}^M \varepsilon_s^2 - b_{\mu\boldsymbol{\theta}}\,,
\end{equation}
where the definition of \( b_{\mu\boldsymbol{\theta}} \) is given in Eq.~\eqref{eq:b_factors2}.

If variations of the likelihood are employed, note that instead of \(N\), the dimension of the NP basis must be used. For instance, in the case of the log-likelihood in Eq.~\eqref{eq:top_comb_log_lik}, one would substitute \((2M_s + M_s^2)\) in place of \((2N + N^2)\).

\section{Generating toy data}
\label{app:C}
To estimate the expectation values of the profile likelihood ratio and the GOF with enhanced precision toy data can be generated by fixing the parameters at their MLE values, \( \hat{\mu} \) and \( \boldsymbol{\hat{\theta}} \). When using the GVM likelihoods, it is also necessary to fix the parameters \( \sigma_{u_s}^2 \) to some value. While one could use the MLE of \( \sigma_{u_s}^2 \), this estimator is biased, and for large \( \varepsilon_s^2 \), it may lead to imprecise results. Instead, it is preferable to use an unbiased estimator at order \( \varepsilon_s^2 \). For Eq.~\eqref{eq:gvm_avareges_full}, the unbiased estimator is defined as
\begin{equation}
    \widehat{\sigma_{u_s}^2}^\ast = \frac{v_s+2\varepsilon_s^2(u_s-\hat{\theta}_s)^2}{1+2\varepsilon_s^2} + 2\varepsilon_s^2 j^{\theta_s \theta_s}\,,
\end{equation}
where \( j^{\theta_s \theta_s} \) is computed as in Eq.~\eqref{eq:b_factors1}. For Eq.~\eqref{eq:prof_comb_log_lik_u_full}, the unbiased estimator becomes:
\begin{equation}
\begin{split}
     \widehat{\sigma_{u_s}^2}^\ast &=\frac{v_s+2\varepsilon_s^2\sum_{i,j}^N(u_s^i-\hat{\theta}_s^i)\left(\rho^{(s)}\right)_{ij}^{-1}(u_s^j-\hat{\theta}_s^j)}{1+2N\varepsilon_s^2} \\ & +2\varepsilon_s^2\text{Tr}\left[j^{\boldsymbol{\theta}_s \boldsymbol{\theta}_s}\left(\rho^{(s)}\right)^{-1} \right]\,,
\end{split}
\end{equation}
where $j^{\boldsymbol{\theta}_s \boldsymbol{\theta}_s}$ is computed as in Eq.~\eqref{eq:b_factors2}. These unbiased estimators were computed following the approach described in~\cite{bib:Cox1968, bib:Codeiro1991, bib:Codeiro2014}. They are then used to generate auxiliary measurements $u_s^i$, as well as the stochastic variables \( v_s \), which follow a Gamma distribution:
\begin{equation}
    v_s \sim \frac{\beta_s^{\alpha_s}}{\Gamma(\alpha_s)}v_s^{\alpha_s-1}e^{-\beta_s v_s}\,,
\end{equation}
where \( \alpha_s = \frac{1}{4\varepsilon_s^2} \) and \( \beta_s = \frac{1}{4\sigma^2_{u_s}\varepsilon_s^2} \).

\end{document}